\documentclass{aa}  

\DeclareRobustCommand{\VAN}[3]{#2}
\let\VANthebibliography\thebibliography
\def\thebibliography{\DeclareRobustCommand{\VAN}[3]{##3}\VANthebibliography}

\usepackage{graphicx}
\usepackage[switch]{lineno}
\usepackage{amsmath}
\usepackage{txfonts}
\usepackage{appendix}
\usepackage{url}
\usepackage{hyperref}
\usepackage{xspace}
\usepackage{color}
\hypersetup{colorlinks=true, citecolor=blue, linkcolor=blue}

\newcommand{\spaceAA}{\AA\xspace}
\newcommand{\Halpha}{\hbox{H$\alpha$} $\lambda$ 6563\xspace}
\newcommand{\Hbeta}{\hbox{H$\beta$} $\lambda$ 4861\xspace}

\newcommand{\HII}{\hbox{H\,{\sc ii}}}
\newcommand{\SIIdf}{\hbox{[S\,{\sc ii}]}$\lambda \lambda$ 4069,4076\xspace}
\newcommand{\OIIdf}{\hbox{[O\,{\sc ii}]}$\lambda \lambda$ 7320,7330\xspace}
\newcommand{\OIIIf}{\hbox{[O\,{\sc iii}]}$\lambda$ 4363\xspace}
\newcommand{\SIIIf}{\hbox{[S\,{\sc iii}]}$\lambda$ 6312\xspace}
\newcommand{\SIIId}{\hbox{[S\,{\sc iii}]}$\lambda \lambda$ 9069,9533\xspace}
\newcommand{\OI}{\hbox{[O\,{\sc i}]}$\lambda$ 6300\xspace}
\newcommand{\NII}{\hbox{[N\,{\sc ii}]}$\lambda$ 6584\xspace}
\newcommand{\NIIf}{\hbox{[N\,{\sc ii}]}$\lambda$ 5755\xspace}
\newcommand{\SIId}{\hbox{[S\,{\sc ii}]}$\lambda \lambda$ 6716,6731\xspace}
\newcommand{\OIId}{\hbox{[O\,{\sc ii}]}$\lambda \lambda$ 3726,3729\xspace}
\newcommand{\OIII}{\hbox{[O\,{\sc iii}]}$\lambda$ 5007\xspace}
\newcommand{\FeII}{\hbox{[Fe\,{\sc ii}]}$\lambda$ 4359\xspace}
\newcommand{\ArIV}{\hbox{[Ar\,{\sc iv}]}$\lambda \lambda$ 4711,4740\xspace}

\newcommand{\ha}{\hbox{H$\alpha$}\xspace}
\newcommand{\hb}{\hbox{H$\beta$}\xspace}
\newcommand{\oii}{\hbox{[O\,{\sc ii}]}\xspace}
\newcommand{\oiii}{\hbox{[O\,{\sc iii}]}\xspace}
\newcommand{\nii}{\hbox{[N\,{\sc ii}]}\xspace}
\newcommand{\sii}{\hbox{[S\,{\sc ii}]}\xspace}
\newcommand{\siii}{\hbox{[S\,{\sc iii}]}\xspace}
\newcommand{\feii}{\hbox{[Fe\,{\sc ii}]}\xspace}
\begin{document} 

   \title{SDSS-IV MaNGA: Data-Model Discrepancy in Temperature-sensitive Line Ratios for Star-forming Galaxies}
   \titlerunning{Data-Model Discrepancy in Temperature-sensitive Line Ratios}


   \author{Ziming Peng \inst{1}\thanks{Email: zmpeng@link.cuhk.edu.hk}
          \and Renbin Yan \inst{1,2}\thanks{Email: rbyan@cuhk.edu.hk}
          \and Xihan Ji \inst{3,4}
          \and Zesen Lin \inst{1,2}
          \and Man-Yin Leo Lee \inst{5,1}
          }

   \institute{Department of Physics, The Chinese University of Hong Kong, Shatin, New Territories, Hong Kong SAR, China\\
            \and
            CUHK Shenzhen Research institute, No.10, 2nd Yuexing Road, Nanshan, Shenzhen, China\\
            \and
            Kavli Institute for Cosmology, University of Cambridge, Madingley Road, Cambridge CB3 0HA, UK\\
            \and
            Cavendish Laboratory, University of Cambridge, 19 JJ Thomson Avenue, Cambridge CB3 0HE, UK\\
            \and
            Department of Astronomy, University of California San Diego, 9500 Gilman Drive, La Jolla, CA 92093, USA
            }

   \date{Received XXX; accepted YYY}

 
  \abstract
   {}
   {Gas-phase metallicity is a fundamental parameter that helps constrain the star-forming history and chemical evolution of a galaxy. Measuring electron temperature through auroral-to-strong line ratios is a direct approach to deriving metallicity. However, there is a longstanding discrepancy between metallicity measured through the direct method and that based on the photoionization models. This paper aims to verify and understand the discrepancies.}
   {We bin $\sim$ 1.5 million spaxels from SDSS-IV MaNGA according to metallicity and ionization parameters derived from theoretical strong-line calibrations. We stack the spectra of spaxels within each bin and measure the flux of strong lines and faint auroral lines. Auroral lines for \oii, \sii, \oiii, and \siii are detected in the stacked spectra of most bins, and the \nii auroral line is detected in fewer bins. We apply an empirical method to correct dust attenuation, which makes more realistic corrections for low ionization lines.}
   {We derive electron temperatures for these five ionic species and measure the oxygen and sulfur abundances using the direct method. We present the resulting abundance measurements and compare them with those model-calibrated strong-line abundances. The chemical abundances measured with the direct method are lower than those derived from the photoionization model, with a median of 0.09 dex. This discrepancy is smaller compared to the results based on other theoretical metallicity calibrations previously reported. However, we notice that the direct method could not account for the variation in ionization parameters, indicating that the precise calibration of metallicity using the direct method has yet to be fully realized. We report significant discrepancies between data and the photoionization model, which illustrates that the one-dimensional photoionization model is incapable of representing the complexity of real situations, and cannot predict the increase of auroral-to-strong line ratio of \oii at high metallicity.}
   {}


   \maketitle
%

\section{Introduction}

Metallicity measurements in galaxies are crucial for understanding the baryon cycles in and around galaxies \citep{maiolino2019re,lilly2013gas}. By analyzing the emission-line spectra of ionized gaseous nebulae, we can determine the abundances of various elements, which reflect the cumulative history of metal enrichment in galaxies. Specifically, star-forming (hereafter SF) regions, or \HII\ regions, are the gaseous nebulae ionized by young hot O-type or B-type stars. Typically, the gas-phase metallicity is traced by the abundance of oxygen, but the abundances of nitrogen, sulfur, and other elements help to provide a more complete physical picture of chemical enrichment. 

Several methods can be applied to derive gas-phase metallicity when the optical spectrum is available \citep{osterbrock2006astrophysics,draine2011physics}. One method is to first measure electron temperatures (T$_e$) using the temperature-sensitive auroral-to-strong line ratios. Both the auroral lines and the strong lines are generated by collisional excitation, so they are also called collisionally-excited lines (CELs). Five commonly used such line ratios are: \OIIIf / \OIII, \SIIIf / \SIIId, \NIIf / \NII, \OIIdf / \OIId, and \SIIdf / \SIId. After electron temperatures are computed, ionic abundances can be derived using flux ratios of metal CELs and hydrogen recombination lines such as \hb \citep{perez2017ionized}. This is usually referred to as the direct method. However, the disadvantage of the direct method is that the auroral lines are usually too faint to be observed, about two magnitudes fainter than their nebular line pair \citep{peimbert2017nebular}. This inherent faintness poses difficulties in obtaining accurate line ratios, particularly in cases where the instruments have limitations in resolution or the signal-to-noise (S/N) ratio of the spectrum is insufficient. 

Besides, recombination lines provide an almost temperature-independent method to measure metallicity \citep{peimbert2013densities,esteban2014carbon}, but the optical recombination lines are even fainter than auroral lines, making them difficult to detect. Also, strong metal CELs can trace metallicity. This method can avoid the measurement of weak lines, but the relations between strong line ratios and metallicity are always under debate. There are generally two ways to calibrate these relations: one is to calibrate empirically with the direct method using data of SF regions for which both strong line ratios and electron temperature measurements are available \citep{pilyugin2005oxygen,pilyugin2016new,brazzini2024metallicity}, and another one is to calibrate theoretically using photoionization models where strong line ratios can be derived theoretically for a given set of input parameters, including metallicity, ionization parameter, ionizing spectra, etc \citep{mcgaugh1991h,dors2011analysing,perez2014deriving,morisset2016photoionization}. However, the results produced by these two methods showed significant discrepancies. Calibrations using the empirical method often yield metallicities that are 0.2-0.6 dex lower than those based on photoionization models \citep{kewley2008metallicity,lopez2012eliminating,blanc2015izi}, with the latter closer to the results provided by recombination lines. 

In this paper, we focus on studying the discrepancy between direct-method metallicity and the model-based strong-line metallicity. Past studies using the direct method made the auroral line measurements either in individual \HII\ regions of galaxies (e.g. \citealt{berg2020chaos,rogers2021chaos}), or in stacked spectra of many galaxies or regions of galaxies. For individual galaxies or \HII\ regions, because of the faintness of auroral lines, this approach requires long exposure times except for very metal-poor galaxies.
Such samples are usually biased towards \HII\ regions with higher temperature due to the ease of auroral line detections in them. Alternatively, it can be measured in stacked spectra whose S/N ratio is improved. Previously, people grouped galaxies or spaxels for stacking by their stellar mass and star formation rate (SFR) \citep{liang2007direct,andrews2013mass,khoram2025direct}. This choice of grouping criteria also tends to average regions of different metallicities and make the results dominated by low metallicity regions, which have relatively stronger auroral lines. \cite{curti2017new} presented a new binning method that was based on the ratios between the strong CELs and hydrogen lines, assuming that galaxies with similar line ratios have roughly the same oxygen abundance. This is a better approach. However, the combination of the two line ratios they used, \oii/\hb and \oiii/\hb, does not uniquely determine metallicity. This can be seen by the self-overlapping model grid in a plot of \oii/\hb vs. \oiii/\hb (e.g., Figure 11 of \citealt{dopita2013new}). A given combination of these two line ratios could still mix \HII\ regions with different metallicities due to variations in ionization parameter.

We adopt a method that is similar in spirit to \cite{curti2017new}. Since both metallicity and ionization parameter can vary significantly among \HII\ regions, we consider them as two primary independent variables. In most photoionization models, they are also the two main variables that make up a 2D model grid. Therefore, to study the discrepancy between direct-method metallicity and model-calibrated metallicity, we need to group galaxies or spaxels by both metallicity and ionization parameters, or their proxies, then stack the spectra in each group to measure the auroral-to-strong line ratios. The problem is how to derive metallicity and ionization parameter independently of auroral-to-strong line ratios. We use combinations of strong line ratios to achieve this. Although strong-line-based metallicity or ionization parameter may not be free of systematics, by properly choosing a set of line ratios, we could identify \HII\ regions with similar physical properties. The underlying assumption is that SF regions that are similar to each other in physical properties ought to have similar line ratios for both strong and weak lines.

The combination of line ratios used has to be able to uniquely determine both metallicity and ionization parameter. Furthermore, the line ratios chosen need to be dust-insensitive so that we do not suffer from dust correction uncertainties. For example, \oiii/\oii and \nii/\oii would be a good combination that can uniquely determine metallicity and ionization parameter, but they are sensitive to dust correction. \cite{ji2020constraining,ji2022correlation} have shown that the combinations of \nii/\ha, \sii/\ha, and \oiii/\hb can uniquely determine metallicity and ionization parameter. These line ratios are insensitive to dust correction, given the proximity in wavelength of each line pair. We compare these line ratios with a photoionization model that can best describe the data distribution in this 3D line ratio space to determine the model-based strong-line metallicity and ionization parameter.

We make use of spatially-resolved spectroscopy data with a spatial resolution of 1-2 kpc in thousands of nearby galaxies. We bin all the spaxels according to their strong-line metallicity and ionization parameter, stack the spectra of spaxels within each bin, and then measure the auroral-to-strong line ratios within each bin to derive their direct-method metallicity. Stacking provides an unbiased average estimate for the population being stacked and is free of sample bias resulting from the requirement of detecting auroral lines. 
In this work, we apply an empirical method to do dust attenuation correction, which is more accurate than the traditional extinction correction method. It is worth noting that we are not only comparing the computed metallicities but also attempting to understand the difference between the line ratios predicted by theoretical photoionization models and those from actual observations. 

This paper is structured as follows. In Section \ref{sec:Method}, we introduce the data we use and how we bin and stack the spectra from many spatial pixels (hereafter spaxels), the subtraction of stellar continuum from the spectra, the fitting method of the emission lines, and the error propagation. In Section \ref{sec:DAC}, we present the dust attenuation correction method. Physical properties, such as electron temperatures and ionic abundances of the stacked spectra, are analyzed in Section \ref{sec:result}. In Section \ref{sec:discrepancy}, we compare the observed data with the photoionization model and discuss the potential reasons for the discrepancies. Finally, Section \ref{sec:summary} concludes the paper.

\section{Method}
\label{sec:Method}

\begin{figure}
	\includegraphics[width=\columnwidth]{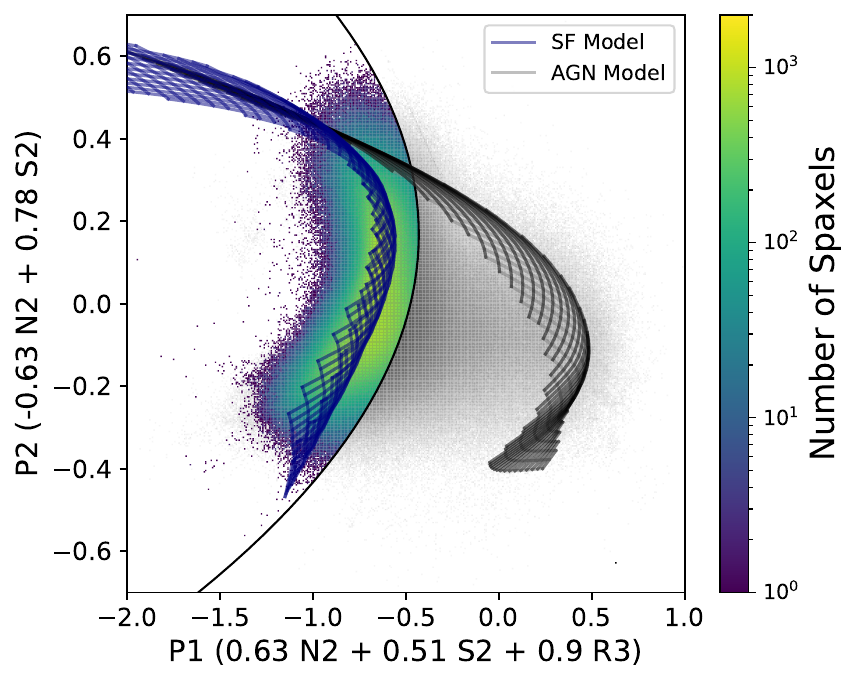}  
    \caption{Density distribution of the data (grey) and the selected samples (colored) in refined optical diagnostic (P$_1$--P$_2$) diagram. The blue grids are the photoionization models for star-forming regions, and the grey grids are the photoionization models for AGNs; each cross corresponds to a unique pair of metallicity and ionization parameter. Both grids are applied a cut so that only the parts that within the middle 98\% of the total data along the hidden P$_3$ axis are shown. The definitions of P$_1$ and P$_2$ are in the text. The demarcation line corresponds to $f_{SF}=0.90$, which means the maximum contamination to \ha from AGN-ionized regions is 10 \%.}
    \label{fig:sf_sample}
\end{figure}

\subsection{Data and Photoionization Model}
\label{sec:2_1}

Our samples come from the Mapping the Nearby Galaxies at Apache Point Observatory (MaNGA) survey \citep{bundy2014overview,yan2016sdss}. We utilize the MaNGA data from the 17th public data release of the Sloan Digital Sky Survey (SDSS, \citealt{abdurro2022seventeenth}). MaNGA utilizes the BOSS spectrographs \citep{smee2013multi} and a multi-object integral field unit (IFU) fiber feed system \citep{drory2015manga} to carry out a survey of more than 10,000 nearby galaxies that covers the optical and near-infrared wavelength range from 3622 to 10,354 \spaceAA. The survey has a spatial resolution of $\sim$ 1 kpc and a spectral resolution of R $\sim$ 2000. The IFU fiber bundles cover a field of view ranging from 12" to 32". The observing strategy is described by \cite{law2015observing} and the flux calibration is described by \cite{yan2016asdss}. The selection of these galaxies is discussed by \cite{wake2017sdss}. The target selections of MaNGA included a primary sample of galaxies covered to a semi-major axis of 1.5$R_e$ and a secondary sample of galaxies covered to 2.5$R_e$ \citep{wake2017sdss}. With the observed data, \cite{law2016data} gives a detailed description of the Data Reduction Pipeline (DRP), and \cite{westfall2019data} gives a detailed description of the Data Analysis Pipeline (DAP), which includes measurements of emission line kinematics and fluxes. \cite{belfiore2019data} provided an analysis on the robustness of these emission line measurements. In this work, we are using emission line fluxes from individual spaxels from DAP (HYB10\_MILESHC\_MASTARHC2). With more than 10,000 spatially resolved galaxies, we obtain an abundant sample not only in terms of the number of galaxies but also in terms of the local distribution of emission lines and local physical properties revealed by each spaxel.

The photoionization model we use in this work for \HII\ regions is generated by the photoionization code CLOUDY v17.03 \citep{ferland20172017}. This model is a simulation for an isobaric \HII\ region with plane-parallel geometry. The ionizing SED is modeled by the software Starburst99 v7.01 \citep{leitherer1999starburst99} with a continuous star-formation history (SFH) of 4 Myr and a Kroupa initial mass function (IMF, \citealt{kroupa2001variation}), and it is computed with the \cite{pauldrach2001radiation} and \cite{hillier1998treatment} stellar atmospheres and a standard Geneva evolutionary track. The hydrogen density of this model is set to 14 cm$^{-3}$, which is derived from the median \sii \ 6716 /\sii \ 6731 of \HII\ regions in MaNGA (\citealt{ji2020constraining}, hereafter JY20). For consistency, each gas-phase abundance is matched with an ionizing SED from stars with the same metallicity. Since the highest stellar metallicity in Starburst99 is only two times solar metallicity, an extrapolation was done to expand the metallicity range to $\sim$ 3.16 times solar metallicity, which is 0.5 in logarithm space. In our photoionization model, it is assumed that other heavy elements scale together with oxygen following the solar abundance pattern, except for secondary elements carbon (C) and nitrogen (N). Secondary C and N are computed using the N/O versus O/H relation derived by \cite{dopita2013new}, where C is fixed to be 1.03 dex more than N before dust depletion. We further include dust depletion using the default depletion factors in CLOUDY \citep{cowie1986high,jenkins1987element}. Detailed descriptions of the photoionization model can be found in JY20.

\subsection{Sample Selection}
\label{sec:sample} 
To ensure that the spectra can have a high S/N for detecting auroral lines, we set an S/N threshold of 5 for H${\alpha}$, H${\beta}$, \NII, \SIId, and \OIII. We discard spaxels with $z\geq 0.08$ as the \siii $\lambda$ 9531 is out of the wavelength range for higher-redshift galaxies. Meanwhile, for the convenience of the stacking procedure, spaxels whose stellar velocity and gas velocity have large offsets, larger than half of the pixel size, which is 34.7 km/s, are excluded. Only 13\% of spaxels have large velocity offsets. After the selection, we check the Equivalent Width (EW) of \ha to ensure most of the spaxels are from SF regions instead of diffuse ionized gas (DIG). According to \cite{lacerda2018diffuse} and \cite{espinosa2020h}, the EW(\ha) threshold of DIG-dominated spaxel is 14 \spaceAA or 6 \spaceAA, lower than which are DIG-dominated spaxels. In our sample, with these two criteria, the fraction of DIG-dominated spaxels is 17.3\% or 0.6\%, respectively. Thus, the result will not be biased by the DIG-dominated spaxels.

For selecting SF regions and discarding active galactic nuclei (AGN), traditionally, the optical diagnostic diagrams proposed by \cite{baldwin1981classification} and \cite{veilleux1987spectral} are widely used (known as BPT/VO diagrams). BPT diagram uses the line ratios of \NII / \ha \ and \OIII / \hb \ to classify the ionization source of spaxels, and the first line ratio can be substituted by \SIId / \ha \ or \OI / \ha. However, these two-dimensional diagnostics are not always consistent with each other, resulting in ambiguous classifications for a significant number of objects (e.g. \citealt{vogt2014galaxy}). In addition, photoionization models of SF regions and AGN ionized regions with varying metallicities and ionization parameters tend to wrap around in BPT diagrams and overlap with each other, implying the demarcation lines defined in 2D might not be able to provide classifications with high purity. Therefore, as a more effective solution for classification, we adopt an optical diagnostic diagram built by the reprojection of the three-dimensional line ratio space as proposed by JY20. Since the reprojection is from a nearly edge-on view of the model surface, this diagnostic diagram simultaneously constrains \NII /\ha, \SIId /\ha, and \OIII /\hb \ line ratios, hence removing the classification inconsistencies between different 2D diagnostic diagrams. Fig. \ref{fig:sf_sample} shows that the reprojection presents nearly edge-on and well-separated photoionization models for SF regions and AGN. Following the demarcation given by JY20, the SF region selection function is given by

\begin{equation}
   P_1<-1.57P_2^2 + 0.53P_2 - 0.48,
	\label{eq:sf}
\end{equation}
where
\begin{equation}
   P_1 = 0.63\,\rm N2+0.51\,S2+0.59\,R3
	\label{eq:p1}
\end{equation}
and
\begin{equation}
   P_2 = -0.63\,\rm N2+0.78\,S2
	\label{eq:p2}
\end{equation}
where N2 = \NII/\Halpha, S2 = \SIId/\Halpha, R3 = \OIII/\Hbeta.

\begin{figure}
	\includegraphics[width=\columnwidth]{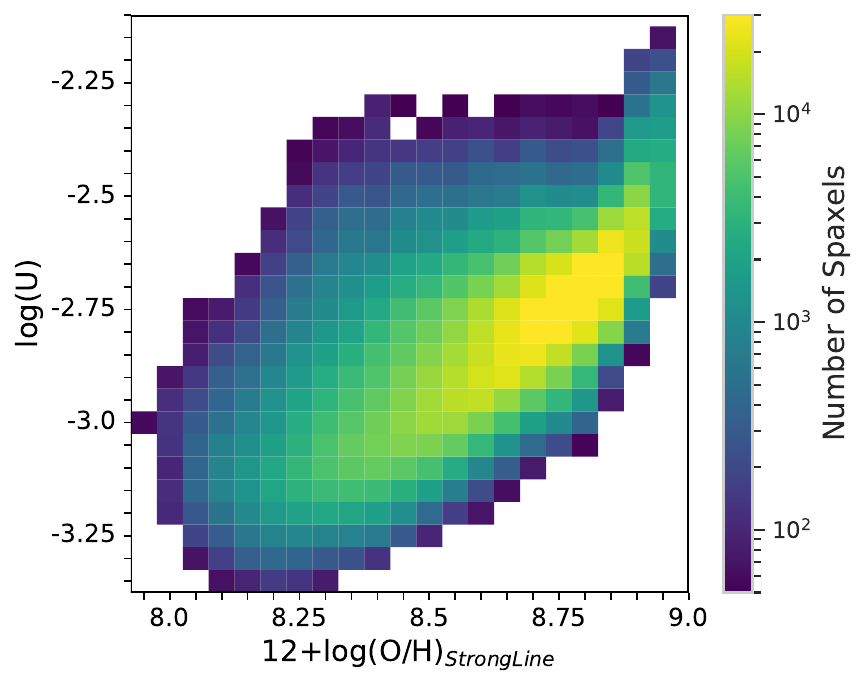}
   \caption{Location of the bins in the metallicity - ionization parameter diagram. Each square represents a bin, and squares are color-coded by the number of spaxels in the bin. Bins with fewer than 50 spaxels are not shown in this diagram.}
   \label{fig:2_2}
\end{figure}

After completing all the selections, $1.45\times 10^6$ spaxels are left in our sample, with a median $z=0.0267$. It should also be noted that we are choosing spaxels simply using the diagnostic diagram, without considering the main ionizing source of the galaxies they originate. 
It illustrates that the focus of this paper is not on discussing the physical properties of SF galaxies, but rather on using the large sample and examining the differences in their emission features and properties after binning.

\begin{table}
   \caption{The statistical differences between median stack and mean stack for strong lines.}               
   \centering          
   \begin{tabular*}{\linewidth}{l c c}     
   \hline\hline       
   Emission line & Median $\Delta$ f/f& 97.5\% $\Delta$ f/f\\
   \hline                    
      \Halpha  & 0.116\% & 0.750\%\\  
      \Hbeta  & 0.128\% & 1.067\%\\ 
      \OIId & 0.120\% & 0.809\%\\ 
      \OIII & 0.126\% & 0.795\%\\
      \SIId & 0.120\% & 0.875\%\\
      \SIIId & 0.146\% & 1.128\%\\
      \NII  & 0.119\% & 0.889\%\\
   \hline
   \end{tabular*}
   \tablefoot{f means the emission line flux measured from the mean stack after the stellar continuum subtraction. $\Delta$ f means the absolute value of line flux measured from the mean stack minus the median stack (both before stellar continuum subtraction).}
   \label{table:2_3}  
   \end{table}

\begin{figure*}
	\centering
	\includegraphics[width=\textwidth]{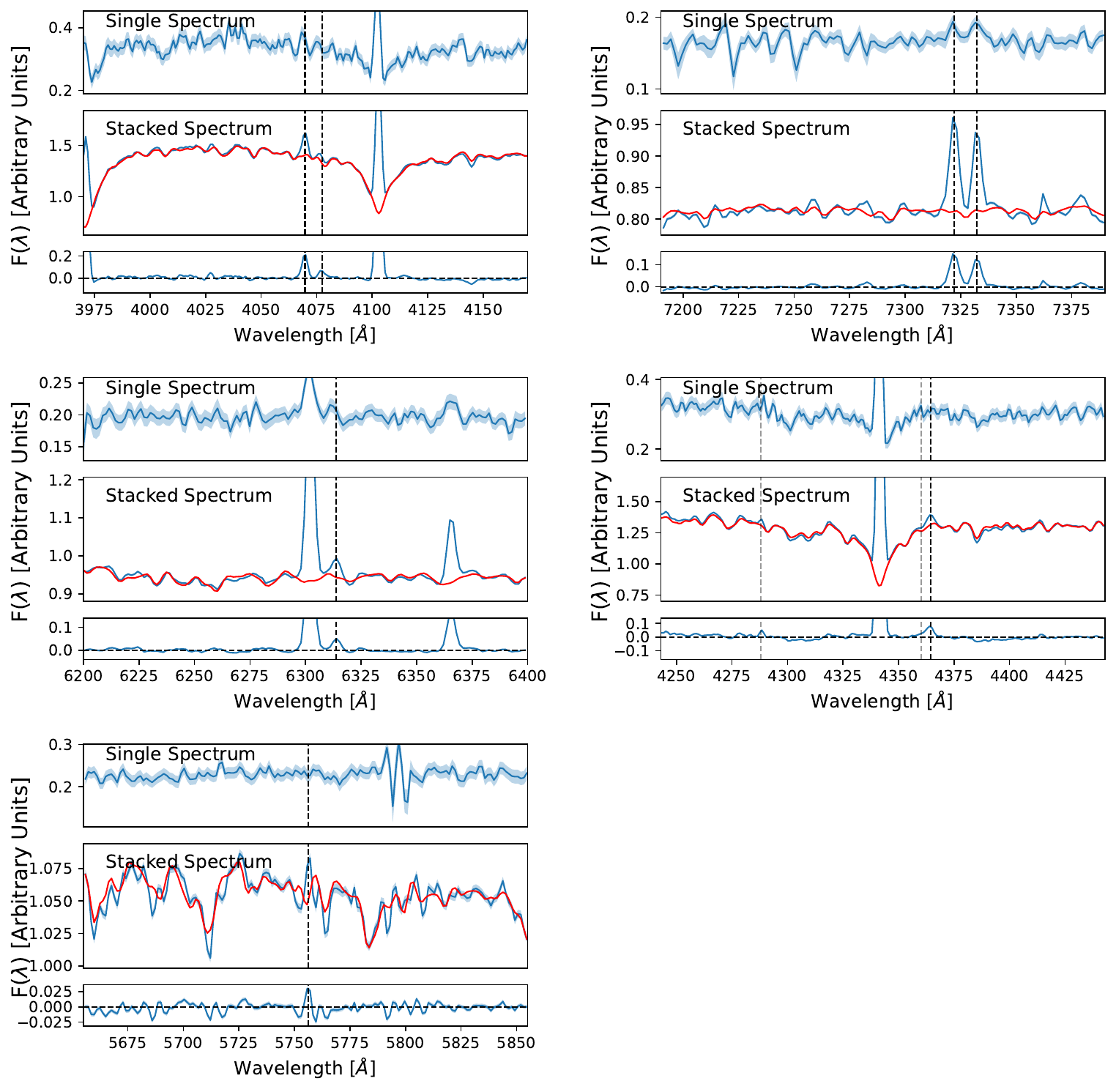}
    \caption{Spectra of \sii auroral lines (top left), \oii auroral lines (top right), \siii auroral line (mid left), \oiii auroral line (mid right), and \nii auroral line (bottom left) from a sample bin (Z$^{-0.10}_{-0.15}$, log(U)$^{-2.80}_{-2.85}$) with 2909 spaxels. In each panel, from top to bottom, the three rows correspond to a single spectrum, the stacked spectrum (blue), as well as the best-fit stellar continuum spectrum(red), and the stacked spectrum after removing the stellar continuum, where the auroral lines can be clearly detected. The shaded regions of the top row represent the uncertainties of flux, taken from MaNGA maps files. The shaded regions of the stacked spectra are errors propagated from individual spaxels. In the \oiii panel, the dashed grey lines represent the \FeII\ (right) and \feii $\lambda$ 4288 (left).}
    \label{fig:2_3}
\end{figure*}

\subsection{Binning and stacking}
\label{sec:stack}
To measure metallicities using the direct method, detecting the faint auroral lines, particularly \OIIIf, is a critical requirement. Because auroral lines can be a hundred times fainter than the strong lines, we enhance the S/N ratio of the spectra by stacking spaxels that are expected to have similar metallicities.

First of all, we estimated the metallicities and ionization parameters of our sample spaxels based on the three strong-line ratios (\nii/\ha, \sii/\ha, \oiii/\hb) using a Bayesian method described by \cite{ji2022correlation}. The ionization parameter, denoted as $U$ and often presented in logarithm ($\log(U)$), represents the relative strength of the ionizing radiation and is defined as $U = \frac{\phi_0}{n_H c}$, where $\phi_0$ is the flux of ionizing photons, $n_H$ is the volume density of hydrogen, and c is the speed of light, ensuring that $U$ is a dimensionless parameter. This parameter serves as an essential tracer of the ionizing state within the spatially resolved region. In this work, we generate CLOUDY models with a metallicity grid sampling the following values in [O/H]: -1.3, -0.7, -0.4, 0, 0.3, and 0.5, and an even ionization parameter grid from -4 to -0.5 with a spacing of 0.5 in log(U). Both the metallicity and ionization parameter grids are interpolated to 100 values. Each metallicity-ionization parameter combination (a node in the model of Fig. \ref{fig:2_2}) will provide a set of line ratios for \nii/\ha, \sii/\ha, and \oiii/\hb according to the emissivities of emission lines. With the input of the line ratios from data, the probability distribution functions (PDFs) could be calculated using Bayesian inference, simultaneously for the metallicity and $\log(U)$. Details for the Bayesian method are described by \cite{blanc2015izi}.

Spaxels with similar metallicities and ionization parameters are likely to correspond to star formation (SF) regions that share comparable properties, such as electron temperature and density. Compared to the commonly used empirical strong-line method (e.g., O3N2 and N2 from \citealt{marino2013o3n2}, N2S2H$\alpha$ from \citealt{dopita2016chemical}), we use the same line ratios as theirs to measure metallicity. However, different from an equation that is straightforward to calculate metallicities, the Bayesian inference doesn't provide an equation but only the probability. The metallicities of the data are derived from their posterior distribution-weighted mean values.

We bin the spaxels based on these Bayesian method-derived metallicities and ionization parameters, using a bin size of 0.05 $\times$ 0.05 dex. For convenience, we abbreviate the bin with [O/H] between $a$ and $a+0.05$, and log(U) between $b$ and $b+0.05$ as (Z$^{a+0.05}_{a}$, log(U)$^{b+0.05}_{b}$) in this work. Figure \ref{fig:2_2} illustrates the number of spaxels within each bin, and we only focus on bins containing more than 50 spaxels. We suspect that bins with an insufficient number of spaxels may not yield a reliable measurement of auroral lines, even after stacking. 

To carry out stacking, we pull out each individual spectrum and its uncertainties from the MaNGA DRP LOGCUBE files. Then we correct for its reddening effect caused by the Milky Way, assuming the reddening map of \cite{schlegel1998maps} and the reddening law following \cite{o1994rnu}. Since different galaxies have different redshifts, and there are velocity offsets for different spaxels in the same galaxies (which are both provided by MaNGA DAP), we shift the spectra back to a logarithmic wavelength grid in the rest frame following the rebinning process described by \cite{lee2024ionized}. In this process, the flux is proportionally allocated to new pixels in the wavelength axis, conserving the flux.

Finally, we normalize each spectrum by the mean flux between 6000-6100 \spaceAA , a relatively feature-free region, and average all the spectra in the same bin to construct the stacked spectrum. The effect of normalization and stacking on line flux ratios can be expressed as 

\begin{equation}
   \frac{f_{\lambda_1}}{f_{\lambda_2}} = \frac{\Sigma_i \frac{f_{i\lambda_1}}{C_i}}{\Sigma_i \frac{f_{i\lambda_2}}{C_i}} = \frac{\Sigma_i \frac{f_{i\lambda_1}}{f_{i\lambda_2}} \frac{f_{i\lambda_2}}{C_i}}{\Sigma_i \frac{f_{i\lambda_2}}{C_i}},
\end{equation}
where $f_{\lambda_1}$ and $f_{\lambda_2}$ are observed flux from wavelength $\lambda_1$ and $\lambda_2$ in a normalized spectrum, $i$ is the $i^{th}$ spaxel stacked, and $C$ is the "weight" for normalization. We also test using the median value instead of the mean as the stacking results. Table \ref{table:2_3} shows the difference in the fluxes of strong lines between median stacks and mean stacks. For strong lines, the differences are around 1\% even for those bins with the largest deviations. For auroral lines, all the residuals of median stacks minus mean stacks are smaller than the uncertainties. We believe that using mean stack or median stack would yield equivalent results for this study.

   

\begin{figure*}
   \centering
   \includegraphics[width=\textwidth]{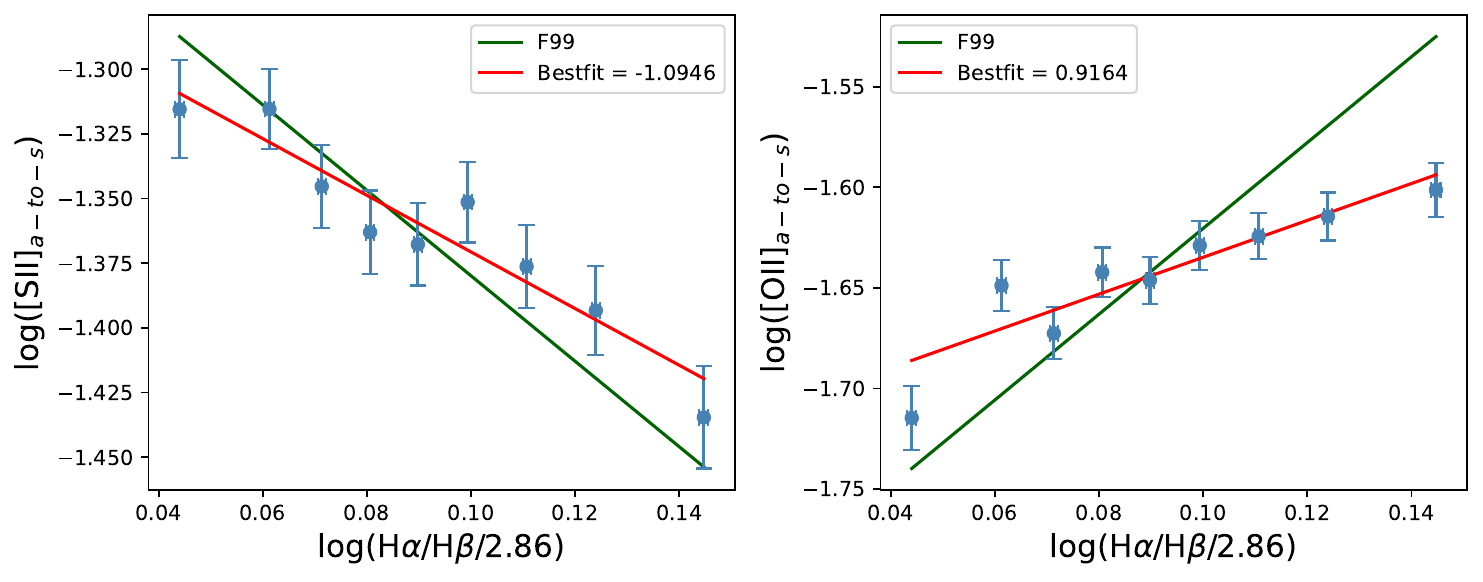}
      \caption{Auroral-to-strong line ratio of \sii \ (left) and \oii\ (right) versus normalized Balmer decrement (\ha\ / \hb\ / 2.86) in one randomly-chosen metallicity-ionization parameter bin, both in logarithm space. Each data point represents the line ratio of a sub-bin that is constructed by spaxels with similar Balmer decrements. The error bars describe the uncertainties of line ratios. The green line is the F99 correction relation with $R_v$ = 3.1, which is generally steeper than the data distribution. The red line is our fitting for these data points, which considers both the x and y axes uncertainties simultaneously.}
         \label{fig:3_1}
\end{figure*}
\subsection{Stellar Continuum Subtraction}
\label{sec:continuum}
After getting high S/N spectra, the stellar continuum needs to be subtracted to measure the emission lines. In this work, we use the stellar templates from MaStar Hierarchical Clustering (hereafter MaStarHC, described by \citealt{abdurro2022seventeenth}). MaStarHC was generated from MaStar \citep{yan2019sdss}, which is a well-calibrated empirical stellar library that used the same set of spectrographs as the MaNGA survey. Thus, the templates have the same wavelength coverage and a similar spectral resolution as MaNGA (R $\sim$ 1800). To achieve better stellar continuum subtraction, we apply the penalized pixel-fitting (pPXF, \citealt{Cappellari2023}) to the stacked spectra. By fitting multiple templates to an observed spectrum, pPXF allows for the robust extraction of stellar kinematics and stellar population properties. 

We follow the continuum modeling and line-fitting procedure adopted by MaNGA DAP, which separates the determination of stellar kinematics from stellar continuum and emission lines fitting. First, we conduct a fitting for the stacked spectrum, only using the stellar templates with an additive Legendre polynomial to determine stellar kinematics.  
In Section 7 of \cite{westfall2019data}, they discussed the influence of different orders of polynomials. We follow their conclusions and use the eighth order. As an additive Legendre polynomial will change the depth of absorption or emission feature, this result is not recommended for use as the stellar continuum of the spectrum. However, it can produce reliable measurements for stellar kinematics, both the velocity offset between the spectra and the templates and the velocity dispersion of the spectra. Masking out regions with emission lines, we do the fitting and mask the pixels with residuals exceeding 3 $\sigma$. Then we iterate the above procedure to convergence and get the stellar kinematics. Fixing the stellar kinematics inputs, we do another fitting using both the stellar templates and the Gaussian emission line templates. In this fitting, the input spectrum is still the stacked spectrum. For emission lines, we fixed the velocity offset of all the lines. However, their amplitudes are free. There are exceptions for several sets of doublets, where their line ratios are fixed. We test the effect of using different orders of multiplicative polynomials and set an eighth-order multiplicative Legendre polynomial for this fitting, following the MaNGA DAP. The multiplicative Legendre polynomials can provide precise subtraction by mimicking the impact of dust effect on stellar continuum \citep{cappellari2017improving}.

Fig. \ref{fig:2_3} demonstrates our stacking and stellar continuum subtraction procedures for \SIIdf\ (top-left panel), \OIIdf\ (top-right panel), \SIIIf\ (mid-left panel), \OIIIf\ (mid-right panel), and \NIIf\ (bottom panel) from one bin which has the strong-line-derived [O/H] range of -0.15 $\sim$ -0.10 and the ionization parameter range of -2.85 $\sim$ -2.80. In each panel, the top row shows a representative spectrum from MaNGA, where the weak auroral lines are "hidden" in the noise that are shown as shaded regions. After stacking spectra in this bin (see Section \ref{sec:stack}), the blue line in the middle row gets vastly increased in the S/N ratio and becomes much smoother. Thus, the auroral lines can be detected. For comparison, the red line is the best-fitting stellar continuum component from the second time fitting of pPXF, which successfully subtracted the stellar continuum while leaving the emission lines emitted by ionized gas regions. The bottom row shows the residuals for the stacking spectrum after stellar continuum subtraction. The auroral lines are clearly visible, at which point the processing of the entire spectrum is complete, and we have obtained emission lines with high S/N ratios.

The goodness of fit for the pPXF fitting is indicated by the reduced $\chi^2$ ($\chi^2_{reduced}$), while $\chi^2_{reduced} \simeq 1$ means the fitting uncertainty is reliable. Among the 329 metallicity-ionization parameter bins, we obtain a mean $\chi^2_{reduced}$ of $\sim$ 2.0, which illustrates that the input uncertainties are underestimated. Therefore, we scale the input uncertainties by $\sqrt{\chi^2}$.

\subsection{Emission Line Flux Measurements}
\label{sec:emline}
We use a similar fitting method as \cite{yan2018shocks} to fit the emission lines for the continuum-removed stacked spectrum. For the singlet, we simultaneously fit a linear continuum from two sidebands near the emission line as well as a single Gaussian profile for the emission line, which has a fixed center on its rest-frame wavelength. Subtracting the linear continuum, we can get rid of spectrum skewing due to the potentially limited continuum-fitting quality. When fitting an emission line, we select the Gaussian flux above the linear continuum as the flux measurement of the emission line, as well as choose the uncertainty of the Gaussian fitting, which originates from the uncertainty of stacked spectra, as the uncertainty. For the doublet, we fit two Gaussian profiles and constrain their wavelength centers by the relatively strong line. We also fix the velocity dispersions of the two lines to be the same.

While dealing with the \OIIIf\ measurement, we notice an emission line near 4359 \spaceAA, which makes a double-peak-like feature near \OIIIf. Similar situations were reported in other papers \citep{curti2017new,arellano2020t}, and it is suspected to be the \feii $\lambda$ 4359. After verification, we find another \feii$\lambda$ 4288 line with the same upper energy level as \feii$\lambda$ 4359, which can also be detected in the stacked spectra. In Fig. \ref{fig:2_3}, they are indicated by grey vertical lines. We are uncertain what affects the flux ratio of \feii lines, but as the metallicity grows, both \feii lines become stronger. It is suggested that there is a fixed line ratio between \feii$\lambda$ 4288 and \FeII; however, in the stacked spectra, we do not find the relation between these two line fluxes.
\begin{table*}
   \caption{The statistical properties of relative attenuation.}             
   \label{table:3_2}      
   \centering          
   \begin{tabular}{l c c c}     
   \hline\hline       
   Emission line ratio & Median slope & Standard deviation & Slope derived by F99 correction law\\ 
                  & (m'$\pm\sigma$') & $\sigma_{std}$ & with $R_v$ = 3.1\\
   \hline                    
      \SIIdf / \SIId & -0.743 $\pm$ 0.032 & 2.115 & -1.652\\  
      \OIIdf / \OIId & 1.465 $\pm$ 0.260 & 0.892 & 2.132\\ 
      \SIIIf / \SIIId & -0.696 $\pm$ 0.173 & 2.232 & -0.889\\ 
   \hline
      \SIId / \OIId & 1.159 $\pm$ 0.006 & 0.230 & 1.903\\
   \hline
   \end{tabular}
   \tablefoot{These median slopes, uncertainties, and standard deviations are derived from the 168 metallicity-ionization parameter bins, which have more than 1000 spaxels, and they are derived in logarithm space of line ratios. We use the 68\% confidence interval of the median slope as the uncertainty ($\sigma$') and also list out the standard deviations of the distributions of the slopes ($\sigma_{std}$). For emission line doublets, F99 corrections use their average wavelength.}
   \end{table*}

\begin{figure*}
   \centering
   \includegraphics[width=\textwidth]{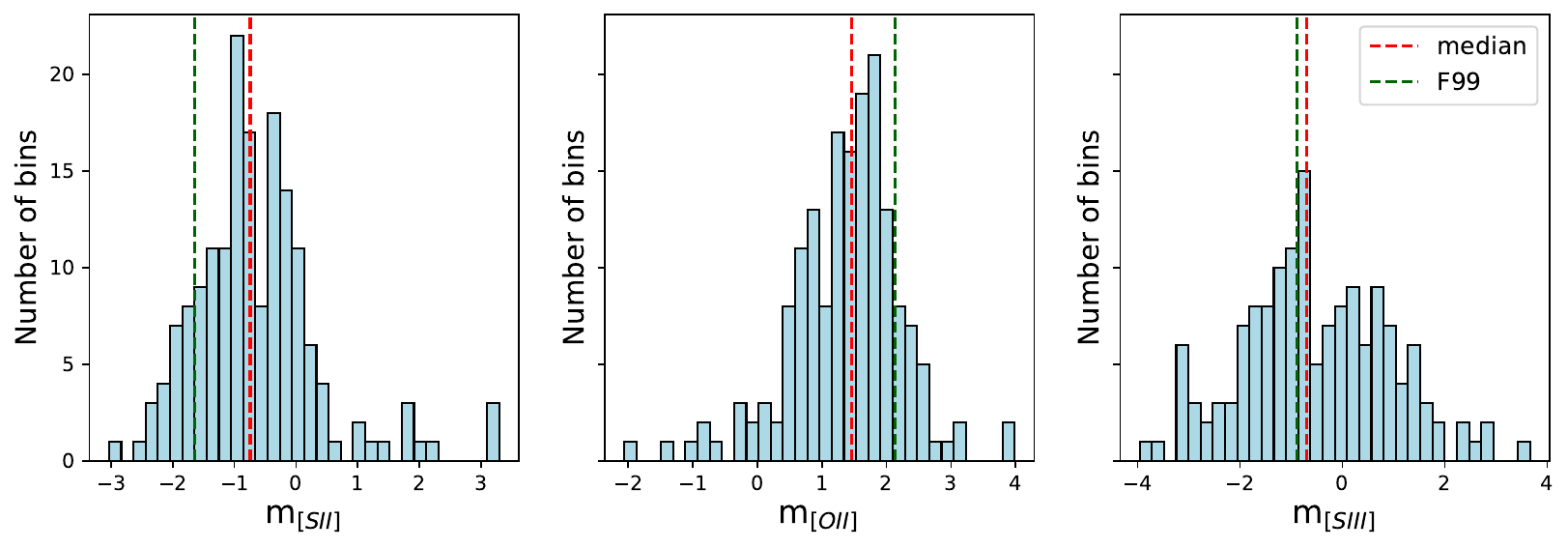}
      \caption{Slopes of relative attenuation distribution in the blue histograms of the linear fitting for \sii$\lambda \lambda$ 4069,4076/ \sii$\lambda \lambda$ 6716,6731 (left) and \oii$\lambda \lambda$ 7320,7330/ \oii$\lambda \lambda$ 3726,3729 (middle), and \siii$\lambda$ 6312/ \siii$\lambda \lambda$ 9068,9531 (right) of all the sub-bins. The green lines represent the slopes of relative attenuation from F99, and the red lines represent the median slopes of 168 bins. The values of median slopes are presented in Table \ref{table:3_2}.}
      \label{fig:3_2}
\end{figure*}
To clear the contamination of \FeII\ line, we use a similar method as \cite{curti2017new}. First, we fit a double-Gaussian profile for \FeII\ and \OIIIf, fixing their velocity width to be the same. It is not necessary for \OIIIf\ and \FeII\ to have the same width; our fixation is only to make sure both emission lines can be fitted. If the best-fitting velocity width is broader than 5 pixels or two wavelength centers are shifted by more than 2 \spaceAA, we can assume \OIIIf\ is undetected. For the other case, similar to \cite{andrews2013mass}, we classified that \OIIIf\ as contaminated if the \FeII\ line flux is $\ge$ 0.5 times of the \OIIIf\ line flux. After two rounds of selection, 228 of the 329 bins are excluded, and the remaining bins are concentrated in sub-solar metallicity regions. We believe that there are many factors other than wavelength that determine the width of emission lines. Therefore, we did not choose to fix the width of \OIIIf\ to H $\gamma$ as done in other works. For the remaining bins, we first subtract the flux of \FeII, then fit a single Gaussian profile for \OIIIf, fixing its center as its wavelength, allowing the changes of amplitude and broadening, to finally get the reliable measurement of \OIIIf\ line flux.  

\section{Dust Attenuation Correction}
\label{sec:DAC}

Dust extinction or attenuation corrections for emission lines are usually done by comparing the observed Balmer decrement with the theoretical Case B value and applying an extinction or attenuation curve (e.g. \citealt{cardelli1989relationship,o1994rnu,fitzpatrick1999correcting,calzetti2000dust}). 
However, this straightforward measurement may not be valid for correcting emission lines in observations with kpc-scale spatial resolution, as \cite{ji2023need} states. Since the spatial resolution of MaNGA is $\sim$ 1-2 kpc, and the typical scale of an SF region is $\sim$ 10-100 pc, the spectra we obtain could be a mixture of SF regions with DIG \citep{zhang2017sdss,mannucci2021diffuse}. And even if we go to smaller scales around 10 pc, we may still see this effect. The issue is that dust extinction is patchy on very small scales, hence different regions with different line ratios can have different extinction as evaluated by an effective E(B-V). As a result, not all emission lines share the same amount of extinction. Also, extinction correction curves, for example \cite{fitzpatrick1999correcting} (hereafter F99), provide a simple extinction correction method, but the actual situations are often more than extinction, so these corrections could be incomplete in explaining attenuation \citep{salim2020dust}. One cannot use a single extinction or attenuation curve to correct all emission lines simultaneously.

As a result of the line ratio differences between DIG and \HII\ regions, \cite{ji2023need} pointed out that low ionization lines have a tendency to have less effective attenuation compared to hydrogen recombination lines, at least for MaNGA data. Therefore, our correction process is questionable if we do not consider the potential inconsistency. So far, all the discussions about dust attenuation correction are limited to strong nebular lines. In the case of auroral lines, other methods have not been tried because of the difficulty in detecting them. For \sii\ and \oii, because the wavelength intervals between auroral lines and strong lines are large, the dust effect becomes a more serious problem. \cite{malkan1983reddening} explores the \sii\ and \oii\ auroral-to-strong line ratios themselves as tracers of dust. However, it is essential to be aware that even a slight difference in line flux measurements could introduce a significant influence on the measurement of electron temperature and metallicity. Consequently, we aim to carefully apply a more accurate method to conduct attenuation correction for these stacked spectra.

\begin{figure*}
   \centering
   \includegraphics[width=0.9\textwidth]{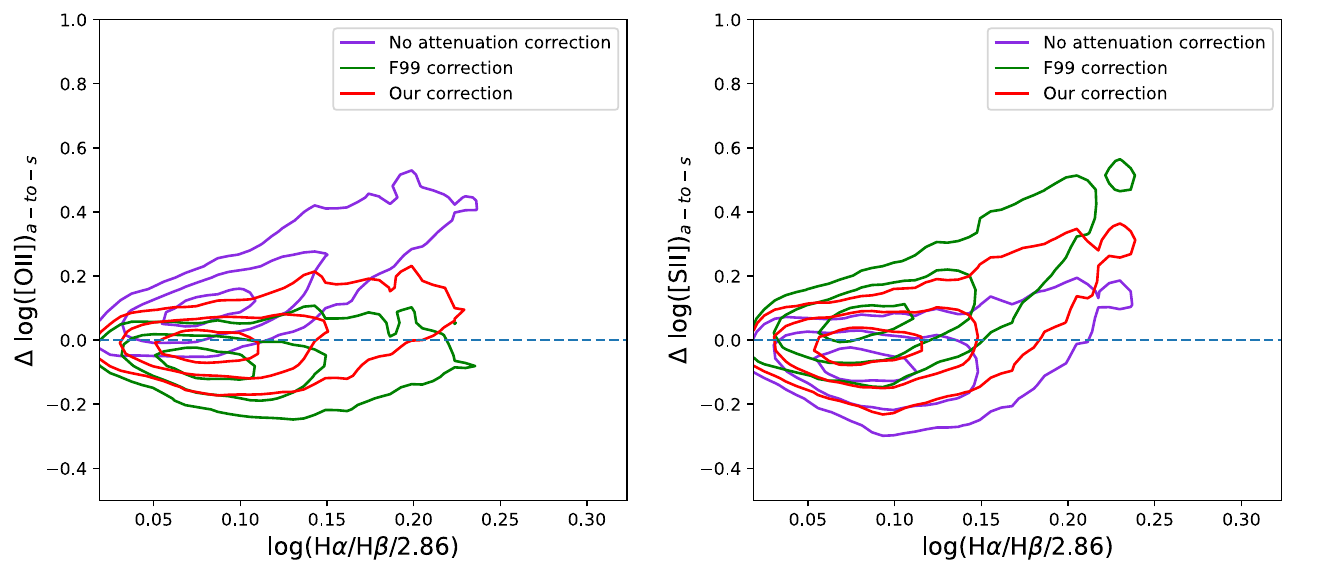} 
      \caption{Relation between \oii(left) and \sii (right) auroral-to-strong line ratios differences and $log(H\alpha/H\beta)$ from all the sub-bins, a total number of 1,512. The line ratio differences mean the actual auroral-to-strong line ratios minus the intrinsic auroral-to-strong line ratios given by the intercept of the best-fit line in each bin. The purple contours represent the density distribution of data without any correction. The green contours represent data after a F99 correction with $R_V$ = 3.1. The red contours represent data after our correcting method described in this section. The contour levels represent 68\%, 95\%, and 99\% of all the data points. The dashed blue lines are horizontal. A horizontal distribution is expected for the correct attenuation correction, as there should be no residual dependence on the Balmer decrement. }
         \label{fig:3_3}
\end{figure*}
Above all, we want to test whether the low-ionization auroral-to-strong line ratios have the same amount of effective attenuation as hydrogen lines. If they have the same tendency as that for strong line ratios, we would like to figure out how different their attenuations are. 
To conduct the test, we divide each metallicity-ionization parameter bin into 9 sub-bins by the \ha / \hb\ value of each spaxel from MaNGA DAP, with each sub-bin having 1/10 of the number of spaxels of the bin, discarding the first and last 5\% to avoid outliers causing offsets by extreme cases. Then, we stack the spaxels for every sub-bin and subtract their stellar continuum, following the procedure in Sec. \ref{sec:stack}. We need to have more rigorous spaxel numbers cut to obtain the S/N ratio of the stacked spectra that enable us to detect faint auroral lines in sub-bins. Here, we only choose those bins that contain more than 1000 spaxels so that there are at least 100 spaxels in each sub-bin, and 168 bins satisfy the criterion. We measure the line ratios and uncertainties for \sii, \oii, and \oiii\ auroral-to-strong lines as well as for \ha / \hb\ in every sub-bin. Since all the sub-bins in one bin have similar metallicities and ionization parameters within a range of 0.05 dex, it can be assumed that they have the same physical properties and intrinsic line ratios. Under these circumstances, the line ratio difference can be attributed to the difference in Balmer decrement, which is expressed as 

\begin{equation}
   \begin{aligned}
      \rm{log} \frac{\it f_{\lambda_1}}{\it f_{\lambda_2}} &= \rm{log} (\frac{\it f_{\lambda_1}}{f_{\it \lambda_2}})_{intrinsic} - 0.4 (A_{\lambda_1}-A_{\lambda_2})\\
      &= \rm{log} (\frac{\it f_{\lambda_1}}{\it f_{\lambda_2}})_{intrinsic} - \frac{A_{\lambda_1}-A_{\lambda_2}}{A_{H\alpha}-A_{H\beta}} log(\frac{\it f_{H\alpha}/\it f_{H\beta}}{2.86})\\
      &= \rm{log} (\frac{\it f_{\lambda_1}}{\it f_{\lambda_2}})_{intrinsic} - m_{\lambda_1,\lambda_2} log(\frac{\it f_{H\alpha}/\it f_{H\beta}}{2.86}),
   \end{aligned}
   \label{eqs:attenuation}
\end{equation}
where $\lambda_1$ and $\lambda_2$ are the wavelength of emission lines, and $A_\lambda$ is the attenuation at wavelength $\lambda$. We carry out a linear regression fit for each of the 9 data points in one bin, and the slope will be the m in Eqs. \ref{eqs:attenuation} which indicate the attenuation difference of $\lambda_1$ and $\lambda_2$ relative to \ha\ and \hb. 

When fitting, we simultaneously consider the uncertainties of Balmer line ratios and the uncertainties of the auroral-to-strong line ratios. We follow the method of \cite{tremaine2002slope}, which gives a likelihood measurement of:

\begin{equation}
   \begin{aligned}
      \ln \space F_{likelihood,i} &= -0.5\chi^2_i \\
      &= -0.5(y_i-mx_i-b)^2/(m^2 \sigma^2_{x_i}+\sigma^2_{y_i}+\sigma^2_0)
   \end{aligned}
\end{equation}
when carrying out a linear fitting $y=mx+b$. $\sigma_x$, $\sigma_y$, and $\sigma_0$ are the uncertainties of the x-axis, y-axis, and the intrinsic scatter of the y-axis. $\sigma_x$ and $\sigma_y$ are computed by the error propagated from each single spectrum. The intrinsic scatter should be the same for all the data during the fit, and we cannot separate it directly from $\sigma_y$. Firstly, we set $\sigma_0$ = 0 and minimize the $\chi^2$ using the minimize function from SciPy once. After that, we set $\sigma_0$ as an unknown, together with slope and intercept, and set the first-time calculated $\sigma_0$ as the new initial guess. Then, we apply the minimize function again. This method originates from \cite{ji2023need}, but we make an improvement so that no iteration is needed.

In Fig. \ref{fig:3_1}, the fitting results of one bin are shown. The red lines are our fitting results using the method described above. For comparison, we also show the best fit using the F99 slope but allowing the intercept to change, which is shown in green. While we consider the x and y uncertainties simultaneously, the uncertainties of the Balmer decrement are small. It is clearly seen that although there are fairly large uncertainties for auroral-to-strong line ratios, our fits match the data points better. 

After we got all the fitting results, histograms are generated in Fig. \ref{fig:3_2} to demonstrate the distributions of the relative attenuation for \sii, \oii, and \siii. For \oii and \sii, it is clear that, although there are large scatters, the slope distributions of the sub-bins deviate from the slopes calculated from F99 (green dashed line). Using the F99 curve with the same E(B-V) as derived from Balmer decrement would overcorrect the attenuation for these two ion species in our data. This is consistent with the trends displayed by the low-ionization strong line ratios shown by \cite{ji2023need}. 
For \siii (right panel of Fig. \ref{fig:3_2}), the median slope is similar to that of F99, also consistent with the trends displayed by high-ionization strong line ratios shown by \cite{ji2023need}. The cause of the different slopes could either be that the extinction curve needs modification, or the attenuations seen by different ions are different. As argued by \cite{ji2023need}, the latter is a much more likely interpretation. 

For every bin, including those that have fewer than 1000 spaxels, we apply this median slope correction to correct the auroral-to-strong line ratios for \oii and \sii. The best-fit ratios at \ha/\hb=2.86 are taken as the intrinsic, unattenuated ratios. Different metallicty-ionization parameter bins would have different intrinsic ratios. We subtract the intrinsic ratios from the observed ratios, merge all bins together, and plot them against \ha/\hb to check for the dependence on Balmer decrement, before and after dust correction. This is shown in Fig. \ref{fig:3_3}, where purple contours denote no corrections, green contours denote F99 corrections, and red contours denote our corrections. Three levels mean 68\%, 95\%, and 99\% of the data points. It is evident that F99 corrections will overcorrect the attenuation by dust and leave an opposite residual dependence, while our correction yields a close-to-zero residual dependence. 

It's possible that the relative levels of attenuation for low-ionization ions compared to hydrogen may have a dependence on ionization parameter and metallicity. We also inspect the metallicity and ionization parameters dependence of slopes for \sii, \oii, and \siii auroral-to-strong line ratios, but they have large uncertainties, and no significant dependence has been found. We further inspect the strong line ratio \SIId / \OIId, whose median slope, uncertainties, and standard deviation are also shown in Table \ref{table:3_2}. As a strong line ratio, \SIId / \OIId\ exhibits a small standard deviation, and all of the slopes are shallower than the F99 correction slope. Plotting the dependence of \SIId / \OIId slopes with two binning parameters, we find that the relative attenuations have a strong correlation with the ionization parameters but are less correlated with metallicities, surprisingly. This suggests that high ionization parameter regions are dustier, and we may need to consider the correlation between relative attenuation and ionization parameter when correcting for dust for low ionization auroral lines.

For \OIIIf\ which represents high ionization zones, due to its faintness in most bins and the minor difference in wavelength for the auroral line and the strong line, we simply apply the F99 correction, assuming this line ratio obeys the same extinction law as Balmer lines. For \siii\ lines, which represent intermediate ionization zones, we try the same fitting method. Different from the low ionization lines, the absolute intensities of \siii\ lines appear to be weaker for both the auroral and the strong lines, and the \SIIIf\ is relatively weak and adjacent to \OI. The right panel of Fig. \ref{fig:3_2} shows that the slope displayed by \siii auroral-to-strong line ratio is similar to that of F99, deviating by only 1 sigma. Thus, we also apply F99 to \SIIIf. For \NIIf, as there are few data points to be considered, and the wavelengths of auroral and strong lines are close, we also use F99.

\section{Results}
\label{sec:result}

\begin{figure}
   \centering
    \includegraphics[width=\columnwidth]{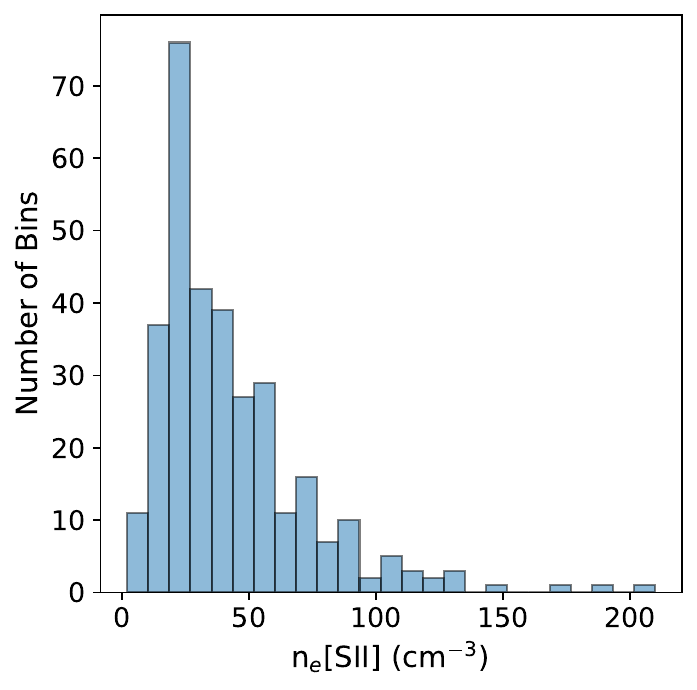}
      \caption{Electron density distribution of all the bins derived from \sii\ 6731 / \sii\ 6716. All the bins are in the low-density limit.}
   \label{fig:4_1}
\end{figure}


\begin{figure*}[!ht]
   \centering
   \includegraphics[width=0.82\textwidth]{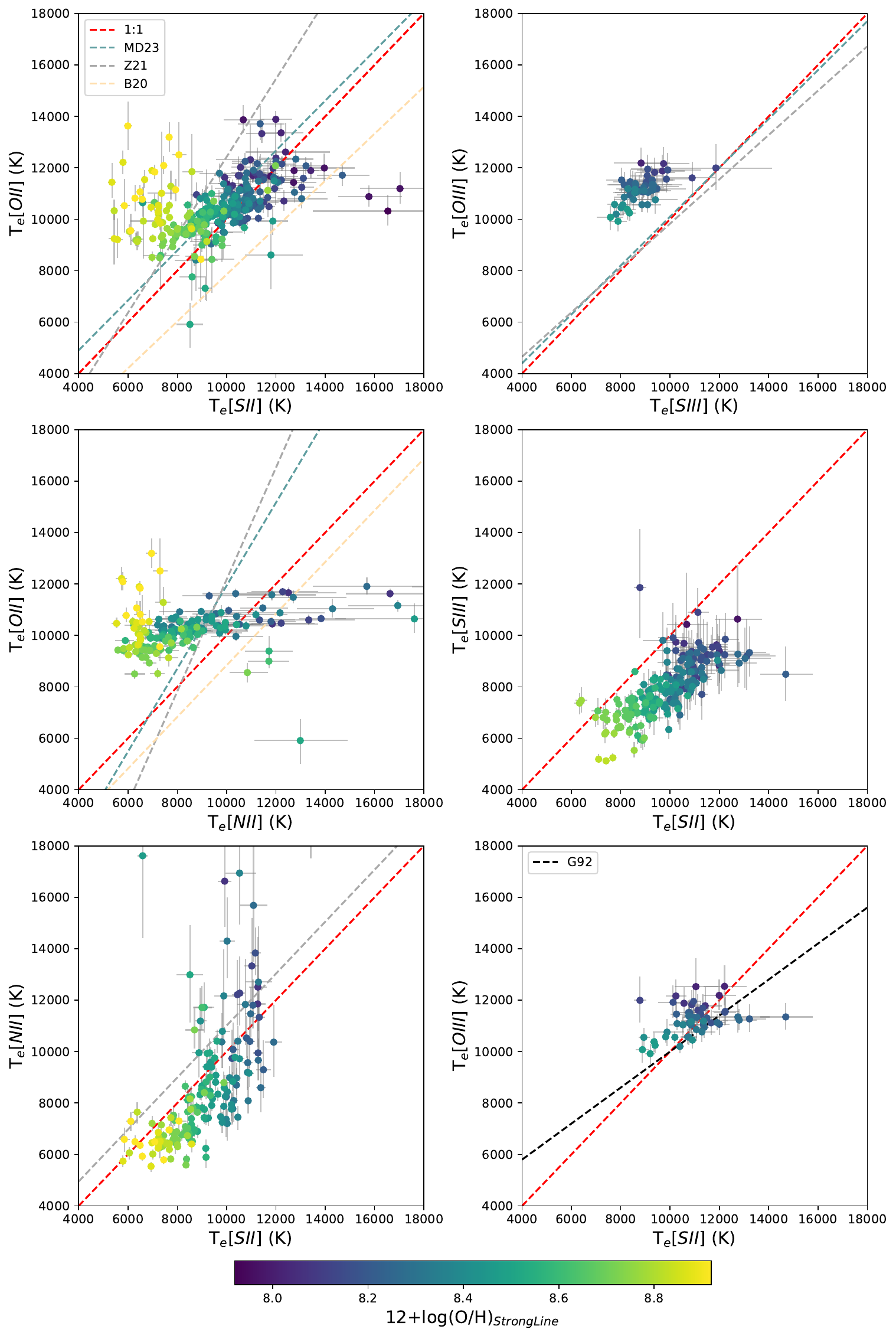}
      \caption{Comparisons between T$_e$ derived from \sii, \oii, \nii, \siii, and \oiii. Each dot represents the data from a stacked spectrum of a certain metallicity-ionization parameter bin. The grey error bars correspond to 1 $\sigma$ uncertainties of the temperature measurements. The red dashed line in each panel shows the 1:1 line. The grey, orange, cyan, and black dashed lines represent previous T-T relations from \cite{zurita2021bar} (Z21), \cite{berg2020chaos} (B20), \cite{mendez2023density} (MD23), and \cite{garnett1992electron} (G92). The color of each dot represents its theoretical strong-line metallicity.}
   \label{fig:4_2}
\end{figure*}

\subsection{Electron Density}
\label{sec:4_1}
We measure the electron temperature using auroral-to-strong line ratios from different ion species representing different ionization zones. The ionization energy of O$^{+}$ is high, so lines from O$^{2+}$ represent regions that are highly ionized. O$^{0}$ and S$^{0}$ have ionization energy close to H$^{0}$ so lines from O$^{+}$ and S$^{+}$ represent low ionization zones. Since S$^{+}$'s ionization energy is between that of H$^{0}$ and O$^{+}$, \siii\ can be a good tracer for intermediate ionization zones. However, in most works applying the direct method to SDSS Legacy data, the intermediate ionization zone is not considered because either the wavelength coverage was insufficient for \SIIId\ or the relative intensity of \SIIIf was much weaker than its neighbor, \OI. Nevertheless, MaNGA enables us to measure both \SIIIf\ and \SIIId\ in this paper.

We use PYNEB version 1.1.18 \citep{luridiana2015pyneb,morisset2020atomic} to estimate the electron temperature and density of the stacked spectrum from each metallicity - ionization parameter bin. The \textit{getCrossTemDen} function in PYNEB enables us to input two sets of line ratios (one density indicator and one temperature indicator) and obtain the electron density and temperature simultaneously based on the five-level atomic structure originating from \cite{de1987five}. For transition probability and collisional strength, we use the newest default atomic data from PYNEB (`PYNEB\_23\_01').

For electron density, we choose the \sii\ 6731 / \sii\ 6716 as the indicator. The doublet has critical densities of $\simeq$ 1400 and 3600 $cm^{-3}$ at $T_e=10^4\ K$, where the collisional and radiative de-excitation reaches an equilibrium. There are other density indicators such as \OIId, \ArIV, even \SIIdf, and \OIIdf, but the latter two are too faint as they are auroral lines, and \ArIV\ is also too faint, although it represents densities in high ionization zones and is worth examining. 
\oii\ 3726 / \oii\ 3729 has a higher critical density than \sii doublet, but we do not use it as the two lines are poorly resolved. In Fig. \ref{fig:4_1}, we present the histogram of densities derived from \sii. Here we assume the electron temperature is $10^4\ K$, since the doublet line ratios are almost independent of temperature when densities are low \citep{nicholls2020estimating}.

\subsection{Electron Temperature}
\label{sec:4_2}
We input the auroral-to-strong line ratio for electron temperature for all five ion species. As the densities typically have values of $n_e$ < 150 $cm^{-3}$, the density variation will not affect their temperature computation for \oii, \nii, and \siii. However, the lower critical densities of \oii\ and \sii\ may lead to a significant difference in temperature measurement with minor changes in density. The uncertainties of temperatures are computed via Monte Carlo simulations. As we know the uncertainties of auroral-to-strong line flux ratio (theoretical error propagation with error computed in Section \ref{sec:emline}), we generate 1000 simulated values of the line ratio following a Gaussian distribution whose $\sigma$ is the line ratio uncertainty for each pair of line ratio in each bin. Among the 1000 temperature measurements we obtain, the 16\% and 84\% values are set as the uncertainty range of the electron temperature.

Fig. \ref{fig:4_2} presents the T$_e$ versus T$_e$ between different ionization species. Focusing on temperatures of low-ionization ions, shown in the three left panels, we first discover that T$_e$ \oii\ follow a different relation with metallicities compared to the other two T$_e$. When the metallicity is low, which corresponds to deep-color dots, T$_e$ \oii\ decreases with the increasing strong-line metallicity of the bin. This is the same as what T$_e$\sii\ and T$_e$\nii\ obey. However, T$_e$\oii\ stops decreasing at high metallicity and shows an upturn trend. We exclude the possibility of improper emission line fitting after checking the spectra of individual bins. Details of this abnormal phenomenon are discussed in detail by Peng et al. (submitted), who argue T$_e$\oii\ is not qualified to represent the low ionization zone due to its offset from the correlation of electron temperature and metallicity. In the top left panel, T$_e$\sii\ and T$_e$\oii\ are relatively consistent and close to the 1:1 line except that they are offset when metallicity is high, and they exhibit a larger dispersion when metallicity is low. However, the data do not stick to the 1:1 line in the middle left and bottom left panels. In both panels, T$_e$\nii\ shows uncertainties more than 3,000 K at lower metallicity, much larger than T$_e$\sii\ and T$_e$\oii. Although T$_e$\nii\ is thought to be a good temperature indicator for low-ionization zones owing to its small scatter and the minor effect of dust on it \citep{berg2015chaos,vaught2024investigating}, T$_e$\nii\ has a larger uncertainty than those of T$_e$\oii\ and T$_e$\sii. In the BOSS spectrographs, the redshifted \NIIf\ happens to be located near the junction between the blue and red cameras, which makes the residual near the emission line noisy (see Fig. \ref{fig:2_2}). Therefore, \NIIf\ is detected in fewer bins than other auroral lines. The large uncertainties of T$_e$\nii\ prevent us from further analyzing \NIIf/\NII. As a result, we can only set the T$_e$\sii\ as the low ionization temperatures.

When it comes to intermediate and high ionization zones, we get fewer data points, mainly concentrating on low metallicity and relatively high ionization parameter bins. T$_e$\oiii\ are uniformly higher than T$_e$\siii, shown in the top right panel. This trend was studied in \cite{binette2012discrepancies}, and they suggested that this may be caused by metallicity inhomogeneity, a $\kappa$-distribution of electron energy, or shock wave. Compared with low ionization lines, T$_e$\oiii\ and T$_e$\siii\ tend to be less dispersed when the binning metallicity and ionization parameter change. Another noteworthy point is that we do not have extremely high T$_e$\oiii\ (typically over 15,000 K), which can probably be attributed to the averaging effect of stacking and better removal of the contamination of \FeII. Comparing T$_e$\siii\ to T$_e$\sii\ in the middle right panel, we find that \sii\ auroral-to-strong line ratios will derive higher temperature than \siii, contrary to other literature and some empirical relations \citep{berg2020chaos,zurita2021bar}. But this can be consistent with the theoretical expectation of the rising temperature in the partially ionized zones. If these two T$_e$ are fitted linearly, the slope will be shallower than 1. For T$_e$\oiii\ versus T$_e$\sii\ in the bottom right panel, there are only a few data points due to the difficulties of detecting \OIIIf\ and getting rid of \feii\ contamination. They are located near the 1:1 line with some scatter.

In this work, we do not aim to derive linear relations for each T$_e$ versus T$_e$, so we only show the 1:1 red dashed lines. We are also not keen on calibrating $t_2$-$t_3$ relations, which provide a reference for the low (high) ionization temperature once high (low) ionization temperature is derived. The limited quality of some data (e.g. \nii) and the unclear physics of electron temperatures (e.g. T$_e$\oii) prevent us from fitting data points linearly. Also, in Fig. \ref{fig:4_2} we only consider the binning metallicity and not the ionization parameter. The results obtained by conducting a linear fit in this case are incomplete. We will utilize the bins that have electron temperature measurements to compute their metallicity.

\begin{figure}
	\includegraphics[width=\columnwidth]{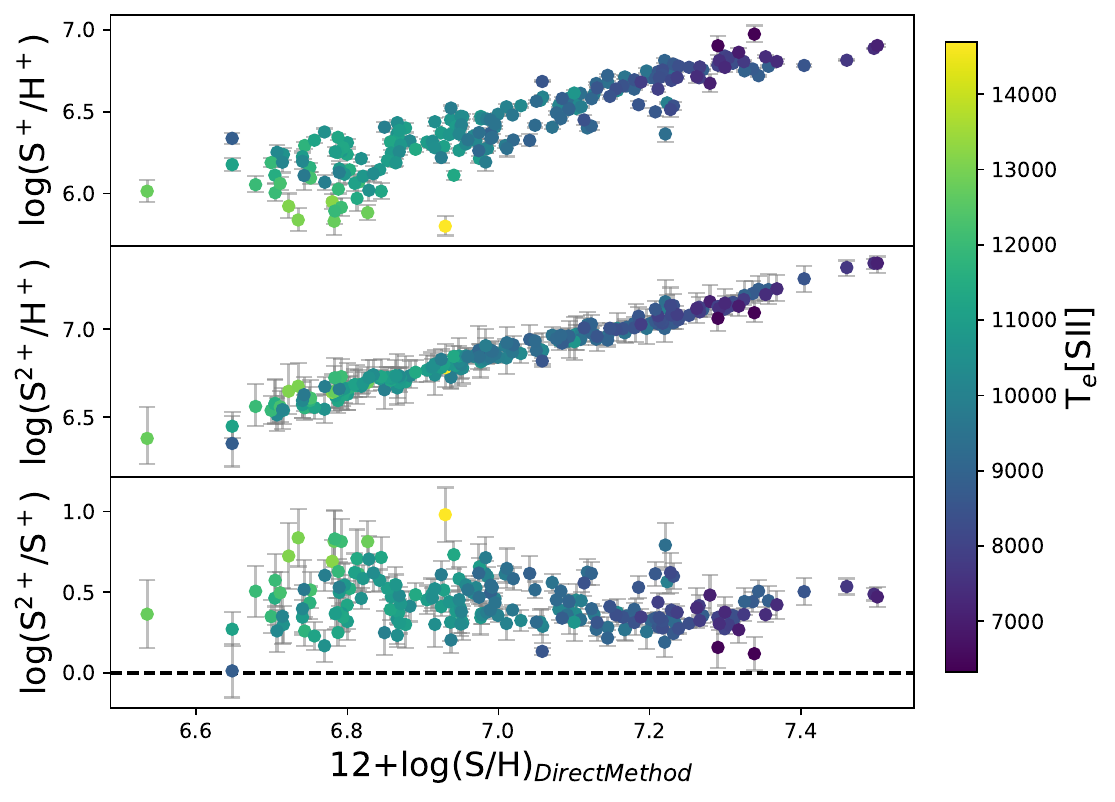}
    \caption{The abundance of S$^+$ (top panel), S$^{2+}$ (middle panel), and S$^{2+}$/S$^+$ (bottom panel) versus total abundance. All the abundances are presented in the form of 12+log(abundance). The uncertainties of abundance measurements are illustrated via the grey error bars, and in the bottom panel, the error bars are calculated via error propagation. The dashed line in the bottom panel demonstrates $12+log(S^{2+}/H) = 12+log(S^+/H) $. Data points are color-coded by T$_e$\sii. } 
    \label{fig:4_3_S}
\end{figure}

\begin{figure}
	\includegraphics[width=\columnwidth]{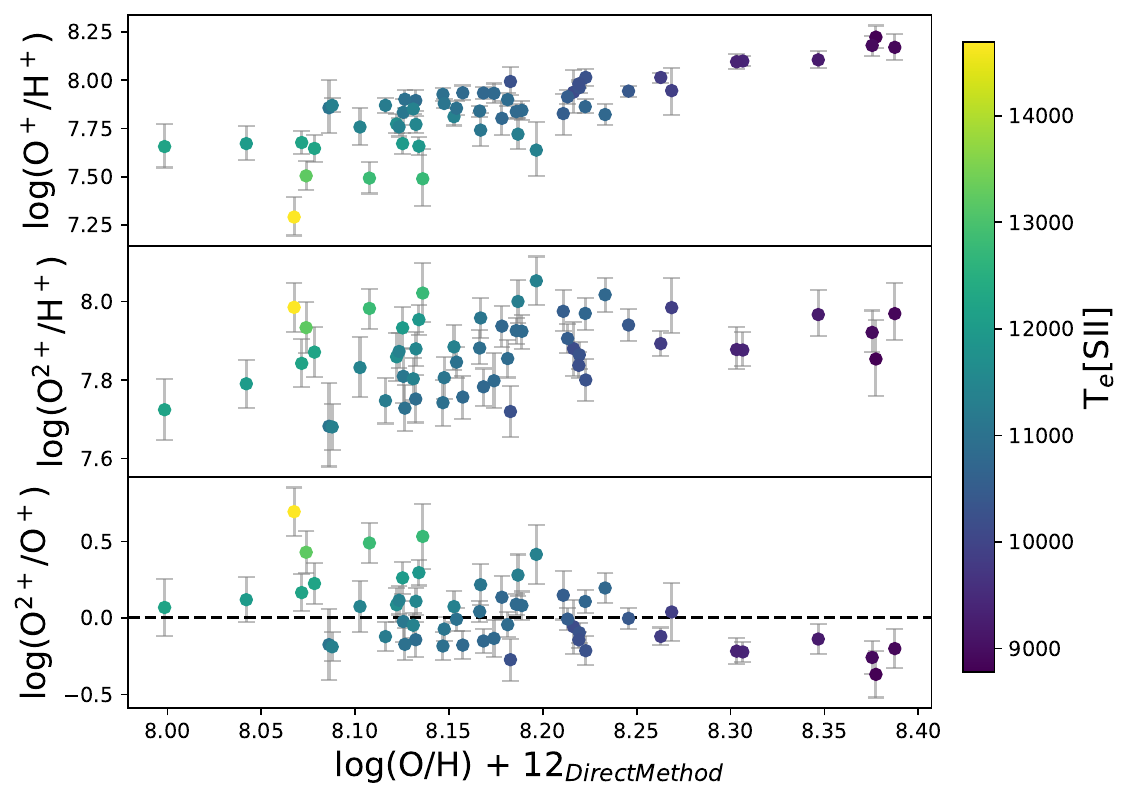}
    \caption{The abundance of O$^+$ (top panel), O$^{2+}$ (middle panel), and O$^{2+}$/O$^+$ (bottom panel) versus total abundance. All the abundances are presented in the form of 12+log(abundance). The uncertainties of abundance measurements are illustrated via the grey error bars, and in the bottom panel, the error bars are calculated via error propagation. The dashed line in the bottom panel demonstrates $12+log(O^{2+}/H) = 12+log(O^+/H) $. Data points are color-coded by T$_e$\sii. }
    \label{fig:4_3_O}
\end{figure}

\begin{figure*}
   \centering
	\includegraphics[width=\textwidth]{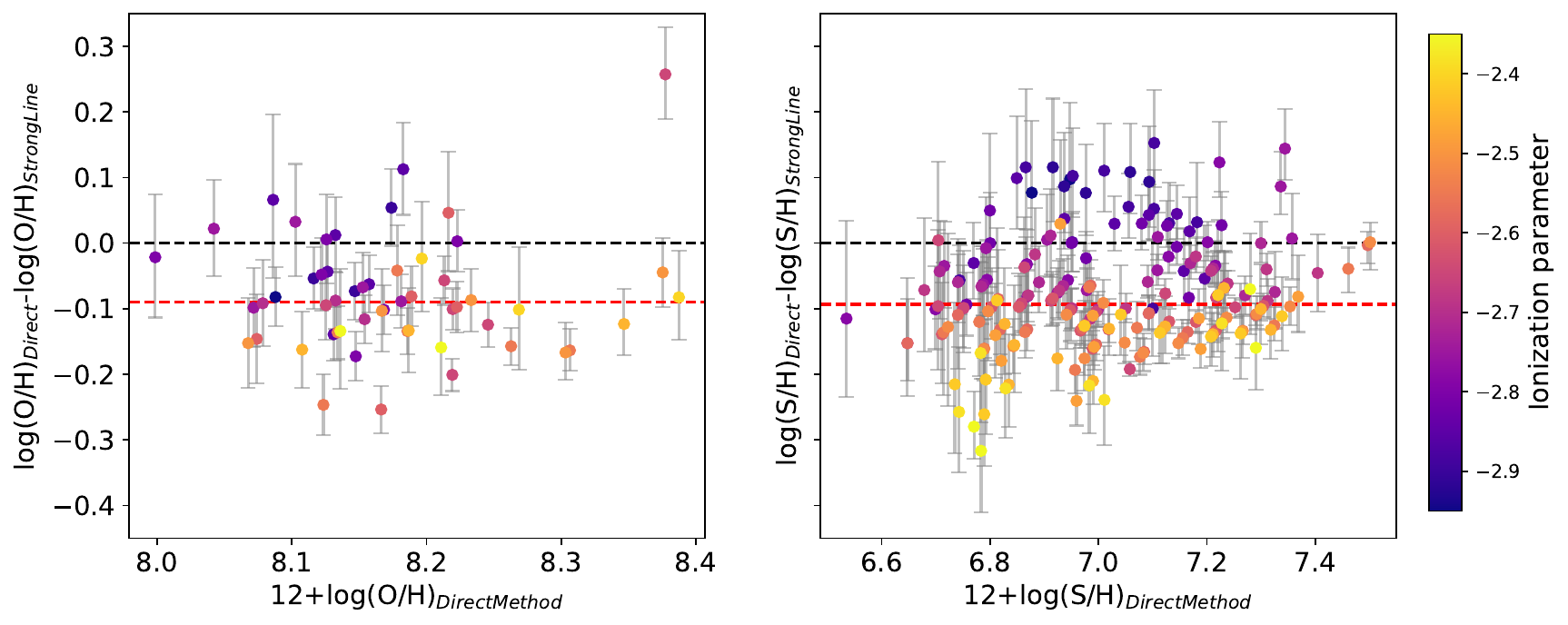}
   \caption{The difference of oxygen (left) and sulfur (right) total abundance derived by the direct method and the theoretical strong-line method versus the total abundance derived by the direct method. The horizontal black dashed lines demonstrate $\rm 12+log(O/H)_{direct} = 12+log(O/H)_{StrongLine}$ and $\rm 12+log(S/H)_{direct} = 12+log(S/H)_{StrongLine}$. The horizontal red dashed lines represent the median values of metallicity differences. The grey error bars show the uncertainties of the direct method. Data points are color-coded by the ionization parameters of their bins.}
   \label{fig:4_3}
\end{figure*}

\subsection{Ionic Abundance}
\label{sec:4_3}
We compute the ionic abundance using the \textit{getIonAbundance} function in PYNEB, which is a Python version of \textit{nebular.ionic}. Since we know the electron temperature, electron density, and nebular line flux relative to the hydrogen recombination lines, for example, \hb, the abundance of a certain ion can be calculated. The uncertainty measurement for ionic abundance is propagated from the uncertainties of electron temperature (See Section \ref{sec:4_2}). We ignore the uncertainty of electron density as the \sii\ doublet are strong lines with very high S/N ratio and the bins are in the low-density limit. 

In the computation of the abundance of an element, applying the ionization correction factor (ICF) is essential for accurately determining total elemental abundance from only two observed ionic ratios. \cite{perez2017ionized} lists the details of how to calculate ICF for each species, and we here apply

\begin{equation}
   ICF(O^+ +O^{2+}) = 1+\frac{y^{2+}}{y^+}
   \label{eq:icf_o} 
\end{equation}
and

\begin{equation}
   ICF(S^+ +S^{2+}) = [1-(\frac{O^{2+}}{O^+ +O^{2+}})^\alpha]^{-1/\alpha},
\end{equation}
where $y^+$ and $y^{2+}$ represent the abundance of He$^+$ and He$^{2+}$. We only apply Eqn. \ref{eq:icf_o} for oxygen if He II 4686 \spaceAA can be detected. Otherwise, we assume that 

\begin{equation}
   \rm \frac{O^+ +O^{2+}}{H^+} = \frac{O}{H}
\end{equation}
and do not apply ICF for oxygen as not enough high-energy ionizing photons are present to produce $\rm He^{2+}$ and $\rm O^{3+}$. For sulfur, the way to calculate the ICF was initially proposed by \cite{stasinska1978empirical}, and we follow \cite{dors2016sulphur} to choose $\alpha$ = 3.27. However, some bins have both S$^+$ and S$^{2+}$ abundance measurements but do not have O$^{2+}$ abundance. We check the ICF for measurable bins, and all of them concentrate in the range from 0.99 to 1.03. We then assume that ICF has an effect of less than 3\% for the final results, and use the default ICF = 1 for those bins that do not have O$^+$ or O$^{2+}$ abundance measurements.


With the correction from ICF, we can derive the total abundance of oxygen and sulfur. 
Fig. \ref{fig:4_3_S} and Fig. \ref{fig:4_3_O} demonstrate the ionic abundance of S$^+$, S$^{2+}$, O$^+$, and O$^{2+}$ across their total abundance we measured. In Fig. \ref{fig:4_3_S}, the scatter plots reveal the relationship between the ionic abundance of S$^+$ and S$^{2+}$, both showing positive correlations consistent with expectations, and S$^{2+}$ having a more linear relation while S$^+$ showing more scatter. In the bottom panel, the S$^{2+}$/S$^+$ appears to be almost independent of total abundance change, while S$^{2+}$ are more abundant than S$^+$. Conversely, Fig. \ref{fig:4_3_O} illustrate the ionic abundances of O$^+$ and O$^{2+}$. For the O$^{2+}$, we apply T$_e$\oiii.Since we do not understand why T$_e$\oii would increase at high metallicity, we also select T$_e$\sii\ to calculate the O$^+$ abundance. O$^+$ abundances are positively correlated with the total abundance, while O$^{2+}$ does not have a significant correlation with the total abundance. As a result, their ratios exhibit a negative correlation with the total abundance. When 12 + log(O/H) > 8.2, most of the bins have more O$^+$ than O$^{2+}$. It is shown in other literature that this ratio will decrease to less than -1.0 at a higher metallicity \citep{curti2017new}. But in our sample, the \oiii\ detection limits our upper boundary of total abundance measured from the direct method. Data points in both figures are color-coded by T$_e$\sii. With the increasing abundances, electron temperatures decrease for both S and O.

We compare the total abundances of oxygen and sulfur derived from our measurements with those obtained using the theoretically calibrated method, which we use to bin the spaxels. Here we use the same assumption as in \cite{ji2022correlation} for the solar abundance of oxygen, that is 12+log(O/H)$_\odot$ = 8.69, originating from \cite{grevesse2011chemical}, and a 0.22 dex deduction due to the depletion into dust. For sulfur, we apply 12+log(S/H)$_\odot$ = 7.15 following \cite{asplund2009chemical}. 
Results for metallicity measurements are shown in Fig. \ref{fig:4_3}. The left panel displays the relationship between the abundance of oxygen (12+log(O/H)$_{\rm direct}$) and the difference between log(O/H) derived from the direct method and their binning metallicities, color-coded by the ionization parameter of each bin. The median value of the difference between direct metallicity and theoretical strong-line metallicity is -0.09 dex. 
The right panel illustrates a similar comparison for sulfur, showing the total abundance of sulfur (12+log(S/H)$_{\rm direct}$) against the differences between the two methods. The median of their difference is also -0.09 dex. In \cite{kewley2008metallicity}, the theoretical-calibrated metallicity and empirical-calibrated metallicity have more than 0.5 dex of difference. Compared with these results, the differences we obtain are relatively small. Both panels are color-coded by the binning ionization parameters, and they exhibit a strong correlation with the total abundance difference. Most of the data points above the horizontal lines are bins with low ionization parameters, and the direct metallicities we obtain are getting smaller compared to the strong-line metallicities with increasing ionization parameters. This phenomenon illustrates that the two methods may carry different systematic errors in their dependence on ionization parameters. 

\begin{figure*}
   \centering
   \includegraphics[width=\textwidth]{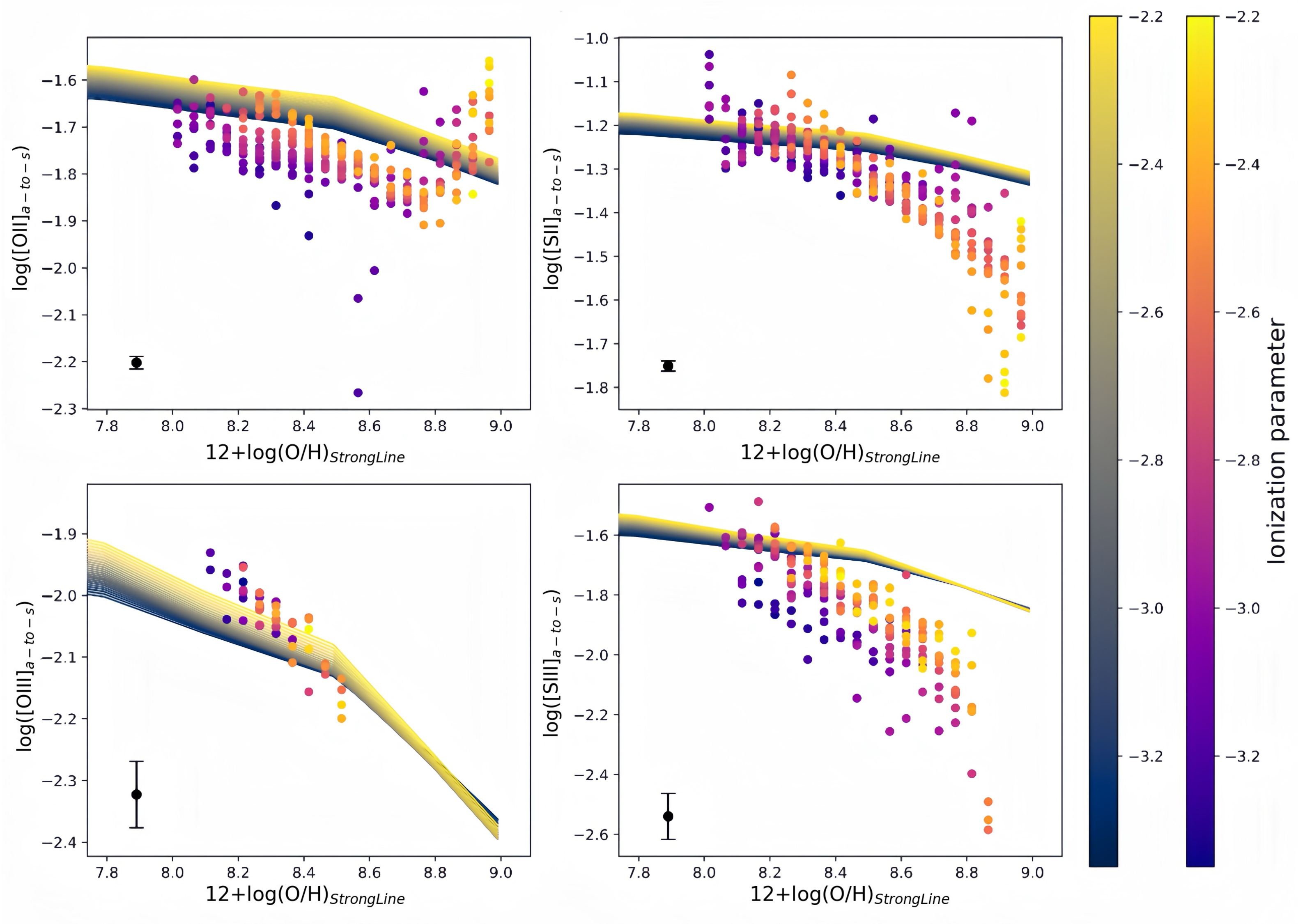}
      \caption{Observed auroral-to-strong line ratios and CLOUDY-generated auroral-to-strong line ratios plotted as functions of binning metallicities for \oii\ (upper-left), \sii\ (upper-right), \oiii\ (lower-left), and \siii\ (lower-right). Lines in the panels represent the data generated by CLOUDY and are color-coded by ionization parameters (the left color bar). Dots in the panels represent the observed data and are also colored by ionization parameters (the right color bar). In the lower left corner of each panel, the black error bar shows the median uncertainty of the observed data. Observed data have a significant offset from the models.}
      \label{fig:5_2}
\end{figure*}


\section{Data-Model Discrepancy}
\label{sec:discrepancy}

The discrepancy between metallicity measured using the direct method and that derived from strong lines based on photoionization models has been known before \citep{lopez2012eliminating,kewley2019understanding}. However, it is unclear which method is more accurate. Both methods have a number of assumptions behind them, and both can be inaccurate. It is important to compare and contrast them directly in observable space, i.e., line ratio space, to see how large the actual discrepancy is. 

The correct photoionization model should be able to simultaneously fit all emission line ratios, including both strong line ratios and auroral-to-strong line ratios. The model we adopt can already simultaneously match the typical \NII /\ha, \SIId /\ha, and \OIII /\hb ratios in most SF regions in MaNGA data. This section will evaluate how well its prediction for auroral-to-strong line ratios matches the data.

\subsection{Compare Data with Model}
\label{sec:5_1}

We compare the observed auroral-to-strong line ratios with the predictions of the photoionization models. The observed data are presented after dust attenuation correction, using F99 for \oiii\ and \siii\ and our specific correction method for \oii\ and \sii. In the comparison, we set the model-based metallicity as the x-axis. Both data and model are color-coded by ionization parameter, with two different color bars. The lines with the left color bar show the model, and the dots with the right color bar show the data.

In Fig. \ref{fig:5_2}, we notice that the data tend to have larger scatter at the same metallicity compared to the model. The auroral-to-strong line ratios for \oiii\ and \siii\ exhibit a steeper downward trend with increasing metallicity compared to the model predictions. 
Except for a few data points, \oiii\ data are located close to the model. However, it is undetected even at slightly sub-solar metallicity. 
\siii\ are almost all below the model. When metallicity is larger than 8.85, they are also undetectable. This is consistent with the commonly seen upper limit of the strong-line method with empirical calibrations, which utilizes metallicity from the direct method. 
For \oii\ and \sii, the observed ratios are roughly consistent with the model in low metallicity regimes, with \oii\ having smaller values and \sii\ exhibiting a steeper downward trend. However, in high metallicity regimes, \oii\ shows a significant upturn, which contradicts model predictions. We do not yet have an explanation for this. In contrast, \sii\ decreases rapidly when the metallicities are high and are distributed far below the model grids. These trends are independent of dust correction. Attributing the upturn of \oii and the steep downturn of \sii to dust attenuation correction would require unrealistically large $\rm R_v$ values. 

Fig. \ref{fig:5_2} also shows the trends of ionization parameters of the data and the model. For the model, auroral-to-strong line ratios slightly increase with increasing ionization parameters at fixed metallicity. Notably, this correlation is reversed for \siii\ and \oiii\ when the model reaches its highest metallicity, which may be due to the extrapolation effect since the ionizing SEDs have an upper limit of 2 times solar metallicity (See Sec. \ref{sec:2_1}). Observed data, however, exhibit a more complicated dependence on ionization parameters. Interestingly, \sii\ demonstrates an inverse dependence on ionization parameters at fixed metallicity. However, the line ratios of the other three species tend to increase with rising ionization parameters. This divergence shows the complexity between ionization states and emission line intensities, suggesting that the current model does not capture all the physical effects in real \HII\ regions.

\subsection{Discussion}
\label{sec:5_2}

There are a number of possible reasons that could explain this inconsistency between the data and the model. First, CLOUDY photoionization model is a one-dimensional model. The assumption of uniform or smooth density distribution is unrealistic. In reality, density fluctuations \citep{mendez2023density}, temperature fluctuations \citep{peimbert1967temperature}, scattering by dust, and contamination by DIG can all make the resulting line ratios different from theoretical predictions. We need more sophisticated 3D modeling incorporating realistic density fluctuations to fully reproduce the observed emission line ratios (e.g. \citealt{jin2023spatially}).  

Not only do the models need to be improved, but there is also a need to obtain better data to test the models. Temperature fluctuation in the SF regions contributes to the auroral-to-strong line ratio measurement. Since the CELs are generally more sensitive to temperature than the hydrogen recombination lines, and the faintness of auroral lines makes them unlikely to be detected in low-temperature zones, the auroral-to-strong line ratios are often biased to a higher value than the average. As a result, the larger fluctuation will result in a significant overestimation of electron temperatures and underestimation of abundances \citep{garnett1992electron,peimbert2017nebular,mendez2023temperature,chen2023accurate}. In the previous studies that used selected \HII\ regions, this effect has a systematic influence, because they may exclude \HII\ regions with lower temperature and result in a low metallicity. 
This has been improved in our stacked spectra, since the stacked spectra yield the average auroral-to-strong line ratio. Still, this effect may bias the temperatures to higher values than the actual average temperature, and result in a slightly lower direct metallicity than the strong-line metallicity.

To better understand the discrepancies, there is still a need for more high-resolution data targeting nearby individual SF regions. Only after careful observation to obtain the distributions of auroral-to-strong line ratios in these SF regions can we gain a deeper understanding of the chemical evolution in these SF regions and what improvements the photoionization models should make. With the next generation of surveys like SDSS-V/LVM \citep{drory2024sdss} and AMASE-P \citep{yan2020prototype,yan2024design}, we will achieve a better comprehension of the auroral-to-strong line ratios from both data and models. There is also a need for more data for high metallicity SF regions where auroral lines are detected. With these data, the real picture of \oii\ and \sii\ can be examined.

\subsection{Comparison to Previous Works}
\label{sec:5-3}

Many reasons lead to the differences in metallicities derived with different strong-line calibrations \citep{stasinska2019can}. For Te-based strong-line methods, calibrations using different data samples and binning methods may give different results, since different data samples may have different biases in sample selection. The results derived from integrated light of galaxies and spatially resolved observations are also not the same. \cite{berg2015chaos} points out that the development of atomic data may also affect the final result of the direct method, hence influence the strong-line calibrations. The line ratios used in strong-line methods may have dependence on physical parameters other than oxygen abundance, for example, log(U) and N/O ratios \citep{kewley2002using,dopita2016chemical}, and this will also cause inconsistencies (see \citealt{kewley2019understanding} for a review). For photoionization model-based strong-line methods, using different photoionization codes (e.g., MAPPINGS, \citealt{sutherland2018mappings}) may give different line ratios. Subtle differences in the input assumptions can likewise lead to discrepant results. Not only do these reasons cause the absolute value of metallicity to be different for the same region, the disparities between methods are also different in different regions. 

There are previous studies that have compared the direct method metallicity with strong-line metallicity. For empirical strong-line methods, some improvements have been made to ensure consistency with direct metallicities, such as revisiting the empirical relation with more data \citep{marino2013o3n2}, or adding more line ratio pairs when deriving the calibration \citep{pilyugin2016new}. \cite{easeman2024optimal} compared several empirical calibrators with sulfur-based direct methods, and it also shows that most of the empirical calibrators deviate from direct methods by $\sim$ 0.2 dex when the 12 + log(O/H) $\sim$ 8.8. For photoionization model-based strong-line methods, \cite{perez2005comparative} compared direct metallicity with several empirical calibrators and one photoionization model calibrator \citep{mcgaugh1991h}, finding that the model-based calibrator has 0.2-0.4 dex difference compared to the paper's new empirical calibrator. Similar discrepancies were also reported in \cite{lopez2012eliminating}. \cite{blanc2015izi} confirmed that model-based strong-line methods are systematically higher than the direct method for 0.2 dex, and \cite{vale_asari_bond_2016} yielded the deviations of 0.2-0.4 dex.
\cite{perez2021extreme} used SDSS single-fiber spectra of a sample of extreme emission line galaxies with 0 < z < 0.49, and the comparison between model-based strong-line metallicity and direct metallicity shows consistency within the uncertainty range (less than 0.1 dex). However, extreme emission line galaxies have different ionizing stellar populations and gas geometry, which makes them not typical in the local universe, and the photoionization models here only cover sub-solar metallicities. The requirement of \OIIIf detection also means the sample may be biased towards those galaxies with a high average electron temperature. Compared with most previous works, the \cite{ji2022correlation} model-based strong-line method used in this paper, although still not perfectly aligned with the direct metallicity, presents a generally smaller deviation from the direct metallicity, even when the metallicity is high ($\sim$ 0.09 dex). Overall, this strong-line method is considered to be a reliable metallicity measurement for MaNGA galaxies. Peng et al. in prep will further discuss the scaling relation derived using this method.

\section{Summary}
\label{sec:summary}

We carry out direct-method metallicity measurements by stacking spectra using IFU datacubes from SDSS-IV MaNGA. We select about 1.5 million star-forming spaxels from SDSS-IV MaNGA DR17 and bin them by their metallicities and ionization parameters derived from the photoionization model described by JY20 using strong lines, and assuming that spaxels with the same metallicity and ionization parameters have the same physical condition and intrinsic emission line ratios. Our conclusions are as follows.
\begin{enumerate}
   \item In most of the bins, \SIIdf\ and \OIIdf\ can be successfully detected. \SIIIf\ can be found in about 2/3 of the bins, and \OIIIf\ can only be detected in sub-solar metallicity bins with relatively high ionization parameters. \FeII\ contaminates \OIIIf significantly. Therefore, we only choose \OIIIf\ measurements after deducting the contamination from bins that can distinguish these two lines. \NIIf\ are detected in a few bins, with large uncertainties because they fall into the wavelength junction of the two channels of the BOSS spectrograph. Using median values to stack produces almost the same result as using mean values to stack.
   
   \item We apply a data-based dust attenuation correction method for correcting the auroral-to-strong line ratios and find that low ionization lines (\sii, \oii) have significantly different attenuation from that of hydrogen derived from Balmer decrement. After our correction, the data do not exhibit residual dependence on the Balmer decrement. For high ionization lines and \nii, using \cite{fitzpatrick1999correcting} extinction curve with the overall attenuation derived from Balmer decrement is sufficient for correcting the reddening effect.
   
   \item Electron temperatures are measured for \sii, \oii, \nii, \siii, and \oiii. T$_e$\siii\ are lower than T$_e$\sii, and T$_e$\oiii\ are similar as T$_e$\sii\ in low metallicity regimes with scatters. 
   
   \item Ionic abundance for S$^+$, S$^{2+}$, O$^+$, and O$^{2+}$ are computed. After considering the ICF, the total abundance of sulfur and oxygen derived from the direct method is systematically lower than the theoretical strong-line metallicity by a median value of -0.09 dex. While still not fully consistent, this deviation is much smaller than the average deviations based on other theoretical metallicity calibrations previously reported. Metallicities derived using this Bayesian strong-line method based on the JY20 model are reliable for MaNGA and MaNGA-like observations with 1-2 kpc spatial resolutions. Scaling relations derived using these more reliable metallicities would be checked in a follow-up paper. 

   \item The discrepancy between direct-method metallicity and the strong-line-derived metallicity gets larger at higher ionization parameter. This could have implications for the redshift evolution studies of the mass-metallicity relations, as samples at different redshifts may be biased differently in their ionization parameters. 
   
   \item Comparing our data to the photoionization model, we have ruled out the possibility that the dust effect could have caused the anomaly of \oii\ and \sii\ auroral-to-strong line ratios when metallicity is high. Observed data present more complex distributions than the model. \sii and \siii auroral-to-strong line ratios decrease faster than the model. \oii auroral-to-strong line ratios exhibit an upturn that does not exist in the model. The observed data show stronger dependence on ionization parameter, and \sii has an opposite dependence on ionization parameter. The one-dimensional photoionization model could not account for all the auroral-to-strong line ratios while fitting all the strong line ratios. 
\end{enumerate}
For future works, we highlight the need to examine the auroral lines in individual SF regions, especially metal-rich SF regions, to better understand the inconsistency of theoretical-calibrated and empirical-calibrated metallicities.

\begin{acknowledgements}
We acknowledge the grant support by the National Natural Science Foundation of China (NSFC; grant No. 12373008) and the support by the Research Grant Council of Hong Kong (Project No. 14302522). ZP thanks Ricky Wai Kiu WONG and Yee Ching LAM for useful discussions. RY acknowledges support by the Hong Kong Global STEM Scholar Scheme (GSP028), by the Hong Kong Jockey Club Charities Trust through the project, JC STEM Lab of Astronomical Instrumentation and Jockey Club Spectroscopy Survey System, and by the NSFC Distinguished Young Scholars Fund (grant No. 1242500304). Z.S.L. acknowledges the support from Hong Kong Innovation and Technology Fund through the Research Talent Hub program (PiH/022/22GS). XJ acknowledges ERC Advanced Grant 695671 ``QUENCH'' and support by the Science and Technology Facilities Council (STFC) and by the UKRI Frontier Research grant RISEandFALL.\\

Funding for the Sloan Digital Sky 
Survey IV has been provided by the 
Alfred P. Sloan Foundation, the U.S. 
Department of Energy Office of 
Science, and the Participating 
Institutions. 
SDSS-IV acknowledges support and 
resources from the Center for High 
Performance Computing  at the 
University of Utah. The SDSS 
website is www.sdss4.org.
SDSS-IV is managed by the 
Astrophysical Research Consortium 
for the Participating Institutions 
of the SDSS Collaboration including 
the Brazilian Participation Group, 
the Carnegie Institution for Science, 
Carnegie Mellon University, Center for 
Astrophysics | Harvard \& 
Smithsonian, the Chilean Participation 
Group, the French Participation Group, 
Instituto de Astrof\'isica de 
Canarias, The Johns Hopkins 
University, Kavli Institute for the 
Physics and Mathematics of the 
Universe (IPMU) / University of 
Tokyo, the Korean Participation Group, 
Lawrence Berkeley National Laboratory, 
Leibniz Institut f\"ur Astrophysik 
Potsdam (AIP),  Max-Planck-Institut 
f\"ur Astronomie (MPIA Heidelberg), 
Max-Planck-Institut f\"ur 
Astrophysik (MPA Garching), 
Max-Planck-Institut f\"ur 
Extraterrestrische Physik (MPE), 
National Astronomical Observatories of 
China, New Mexico State University, 
New York University, University of 
Notre Dame, Observat\'ario 
Nacional / MCTI, The Ohio State 
University, Pennsylvania State 
University, Shanghai 
Astronomical Observatory, United 
Kingdom Participation Group, 
Universidad Nacional Aut\'onoma 
de M\'exico, University of Arizona, 
University of Colorado Boulder, 
University of Oxford, University of 
Portsmouth, University of Utah, 
University of Virginia, University 
of Washington, University of 
Wisconsin, Vanderbilt University, 
and Yale University.\\

\textit{Softwares:} Starburst99 \citep{leitherer1999starburst99}, CLOUDY \citep{ferland20172017}, pPXF \citep{Cappellari2023}, Astropy \citep{Astropy_Collaboration_and_Price-Whelan_The_Astropy_Project_2022}, Numpy \citep{2020NumPy-Array}, Matplotlib \citep{Hunter_Matplotlib_A_2D_2007}, extinction \citep{barbary_extinction_2016}.
\end{acknowledgements}

%
\bibliographystyle{aa} 
\bibliography{aanda} 
%

\end{document}